 \documentclass[eqsecnum,aps,amsfonts,amssymb,epsfig,prb]{revtex4} 
\usepackage{graphicx}
\def\singlespace{ \renewcommand{\baselinestretch}{1} \large\normalsize }

\begin{document}


\def\reff#1{(\ref{#1})}
\newcommand{\be}{\begin{equation}}
\newcommand{\ee}{\end{equation}}
\newcommand{\<}{\langle}
\renewcommand{\>}{\rangle}

\def\spose#1{\hbox to 0pt{#1\hss}}
\def\ltapprox{\mathrel{\spose{\lower 3pt\hbox{$\mathchar"218$}}
 \raise 2.0pt\hbox{$\mathchar"13C$}}}
\def\gtapprox{\mathrel{\spose{\lower 3pt\hbox{$\mathchar"218$}}
 \raise 2.0pt\hbox{$\mathchar"13E$}}}

\def\bsigma{\mbox{\protect\boldmath $\sigma$}}
\def\bpi{\mbox{\protect\boldmath $\pi$}}
\def\smfrac#1#2{{\textstyle\frac{#1}{#2}}}
\def\smhalf{ {\smfrac{1}{2}} }

\newcommand{\re}{\mathop{\rm Re}\nolimits}
\newcommand{\im}{\mathop{\rm Im}\nolimits}
\newcommand{\tr}{\mathop{\rm tr}\nolimits}
\newcommand{\fr}{\frac}
\newcommand{\diti}{\frac{\mathrm{d}^2t}{(2 \pi)^2}}
\newcommand{\CITE}{\cite}

\def\Z{{\mathbb Z}}
\def\R{{\mathbb R}}
\def\C{{\mathbb C}}

\title{Two-parameter model predictions and $\theta$-point crossover 
for linear-polymer solutions}

\singlespace
\author{ Sergio Caracciolo}
\address{Dipartimento di Fisica, Universit\`a degli Studi di Milano,  \\
  and INFN -- Sezione di Milano I \\
  Via Celoria 16, I-20133 Milano, Italy \\  
  e-mail: {\tt Sergio.Caracciolo@mi.infn.it}}
\author{Bortolo Matteo Mognetti}
\address{Institut f\"ur Physik, Johannes Gutenberg-Universit\"at, \\
  Staudinger Weg 7, D-55099 Mainz, Germany \\
  e-mail: {\tt mognetti@uni-mainz.de}} 
\author{ Andrea Pelissetto }
\address{
  Dipartimento di Fisica, Universit\`a degli Studi di Roma ``La Sapienza'' \\
  and INFN -- Sezione di Roma I  \\
  P.le A. Moro 2, I-00185 Roma, Italy \\
  e-mail: {\tt Andrea.Pelissetto@roma1.infn.it }}

\thispagestyle{empty}   


\begin{abstract}

We consider the first few virial coefficients of the osmotic pressure,
the radius of gyration, the hydrodynamic radius, and the end-to-end
distance for a monodisperse polymer solution. We determine the 
corresponding two-parameter model functions which parametrize
the crossover between the good-solvent and the ideal-chain behavior.
These results allow us to predict the osmotic pressure and the polymer size
in the dilute regime in a large temperature region above the 
$\theta$ point.

\bigskip 



\end{abstract}
\maketitle

\clearpage

\section{Introduction}

Polymeric fluids exhibit a rich and complex set of phenomena associated both
with system-specific and global properties of the polymer molecules.
Chemical details become increasingly less relevant for global polymer 
properties as the degree of polymerization $N$ increases.
\cite{deGennes-72,deGennes-79,Freed-87,dCJ-book,Schaefer-99}
Thus, for $N\to\infty$, one can use coarse-grained models in which only the 
most fundamental aspects of the polymer structure are taken into account.
The behavior of polymer solutions depends in general on temperature. 
For $T$ large enough, the most relevant feature is the local repulsion.
In this regime, usually called {\em good-solvent} regime, the radius of 
gyration $R_g$, as well as any other quantity that is related to the global
size of the polymer, scales as $N^\nu$, where $\nu$ is a universal exponent;
$\nu \approx 0.5876$ (Ref.~\CITE{bestnu}). As $T$ is lowered, one reaches the 
$\theta$ temperature, $T_\theta$, below which polymers are compact
($R_g\sim N^{1/3}$) and phase separation occurs.\cite{order} 
At the $\theta$ point 
polymers behave approximately as Gaussian coils. The crossover from
good-solvent to $\theta$ behavior is well understood. For $N\to\infty$
any global quantity ${\cal O}$ behaves as:\cite{Duplantier,Schaefer-99}
\begin{equation}
{\cal O}(T,N,c) = \alpha_1 {\cal O}_G(N,c) 
  f_{\cal O}[\alpha_2 (T-T_\theta) N^{1/2} (\ln N)^{-4/11},
  \alpha_3 c \hat{R}_g^3(T,N)].
\label{scaling-gen}
\end{equation}
Here $c$ is the polymer number density, $\hat{R}_g(T,N)$ is the zero-density
radius of gyration, and ${\cal O}_G(N,c)$ is the expression of ${\cal O}$
for ideal chains. The function $f_{\cal O}(x,y)$ is universal, all chemical 
details being included in the constants $\alpha_i$.

Eq.~(\ref{scaling-gen}) is strictly valid only for $N\to\infty$,
$T\to T_\theta$, at fixed $\alpha_2  (T-T_\theta) N^{1/2} (\ln N)^{-4/11}$.
For finite values of $N$ one should also take into account the corrections
to Eq.~(\ref{scaling-gen}) that decay very slowly, as inverse powers 
of $\ln N$. In Ref.~\CITE{PH-05} we computed the crossover curve 
for the interpenetration ratio $\Psi$ and found that logarithmic 
corrections are only relevant very close to $T_\theta$. Outside 
a {\em tricritical region} around $T_\theta$, 
Eq.~(\ref{scaling-gen}) provides a reasonably accurate description of the 
crossover. As emphasized in Ref.~\CITE{Sokal-94}, in order to compute the 
crossover functions defined in Eq.~(\ref{scaling-gen}) one can use the 
continuum two-parameter model (TPM).\cite{Yamakawa-71} Indeed, if we identify
$(T-T_\theta) N^{1/2} (\ln N)^{-4/11}$ with the Zimm-Stockmayer-Fixman
\cite{ZSF-53} variable $z$ (with a model-dependent proportionality factor),
then the crossover function for $\cal O$  corresponds exactly to its TPM 
expression.

It is interesting to note that the TPM is also of interest to describe the 
corrections to scaling in some polymeric systems. Indeed, 
as discussed in Ref.~\CITE{CCPRV-01}, the TPM describes the approach 
to the scaling limit when $V_m/\l^3\ll 1$, 
where $\l$ is the persistence length
and $V_m$ is the volume occupied by a polymer blob of length $l$.

In this paper we wish to compute the crossover functions for several quantities
whose behavior in the good-solvent regime has been considered in 
Refs.~\CITE{CMP-vir,CMP-raggi}. We compute numerically the TPM predictions
for the second, third, and fourth virial coefficient, for the swelling 
factors, and the density corrections to the radius of gyration, 
the end-to-end distance, and the hydrodynamic radius. This allows us to give 
exact predictions for the thermodynamic behavior and for the polymer size 
in the whole dilute regime $\Phi_p\lesssim 1$, where $\Phi_p$ is the 
polymer volume fraction.

The paper is organized as follows. In Sec.~\ref{sec2} we define the TPM
as the scaling limit of the lattice Domb-Joyce model. This is a rigorous 
well-defined definition that does not rely on perturbative field theory.
In Sec.~\ref{sec3} we define the quantities that are considered in the 
paper and report some results and properties that are useful in the
following section. In Sec.~\ref{sec4} we give the results of our work.
We first report the analysis of the Monte Carlo results and then
determine the TPM functions associated with the different quantities.
In Sec.~\ref{sec4.3} and \ref{sec4.4} we use these results to predict the 
osmotic pressure and the polymer size in the dilute and in the semidilute 
regime. Finally, 
in Sec.~\ref{sec4.5} 
we compare our predictions with the available renormalization-group results.
Some conclusions are presented in Sec.~\ref{sec5}.

\section{The Domb-Joyce model} \label{sec2}

In order to compute the TPM crossover functions, 
we consider the three-dimensional lattice Domb-Joyce (DJ)
model.\cite{DJ-72} We consider a cubic lattice and model a polymer of 
length $N$ as a random walk $\{{\mathbf r}_0,{\mathbf r}_1,\ldots,{\mathbf r}_N\}$ with
$|{\mathbf r}_\alpha-{\mathbf r}_{\alpha+1}|=1$ on a cubic lattice.
To each walk we associate a Boltzmann factor
\begin{equation}
e^{-\beta H} = e^{-w\sigma},\qquad\qquad
\sigma = \sum_{0\le \alpha < \beta \le N}
   \delta_{{\mathbf r}_\alpha,{\mathbf r}_\beta},
\end{equation}
with $w > 0$. The factor $\sigma$ counts how many
self-intersections are present in the walk. This model is similar
to the standard self-avoiding walk (SAW) model, in which polymers are
modelled by random walks in which self-intersections are forbidden.
The SAW model is obtained for $w = +\infty$. For finite positive $w$
self-intersections are possible although energetically penalized.
For any positive $w$, this model has the same scaling limit of the
SAW model\cite{DJ-72} and thus allows us to compute the
universal scaling functions that are relevant for polymer solutions.

The DJ model can be efficiently simulated by using the pivot
algorithm.\cite{Lal,MacDonald,Madras-Sokal,Sokal-95b}
For the SAW an efficient implementation is discussed in
Ref.~\CITE{Kennedy-02}. The extension to the DJ model is straightforward,
the changes in energy being taken into account by means of
a Metropolis test. Such a step should be included carefully
in order not to loose the good scaling behavior of the CPU time
for attempted move. We use here the implementation discussed in
Ref.~\CITE{CPP-94}.

The TPM results can be derived from simulations of the DJ model. Indeed,
the continuum results are obtained\cite{BD-79} by taking the limit 
$w\to 0$, $N\to \infty$ at fixed product $w N^{1/2}$ (we call it $x$).
The variable $x$ interpolates between the ideal-chain limit ($x=0$) and 
the good-solvent limit ($x=\infty$). Indeed, for $w = 0$ the DJ model is 
simply the random-walk model, while for any $w\not=0$ and $N\to \infty$
one always obtains the good-solvent scaling behavior. The variable 
$x$ is directly related to the variable $z$ that is usually used 
in the TPM context:\cite{BD-79} indeed, $z = \alpha x$. The normalization 
factor $\alpha$ can be fixed by considering 
the small-$z$ behavior of the interpenetration 
ratio $\Psi$: conventionally one takes $\Psi = z + O(z^2)$. 
In the DJ model\cite{BN-97} $\Psi \equiv (3/2 \pi)^{3/2} w N^{1/2}$
for small $w N^{1/2}$, so that we can identify
\begin{equation}
  z = \left({3\over 2\pi}\right)^{3/2} w N^{1/2}.
\end{equation}

\section{Definitions} \label{sec3}

We consider the osmotic pressure $\Pi(T,N,c)$ or, equivalently, the 
adimensional compressibility factor
\begin{equation}
Z(T,N,c) \equiv {M \Pi\over RT\rho} = {\Pi\over k_BTc},
\end{equation}
where $c$ is the polymer number density, $\rho$ the weight concentration,
$M$ the molar mass of the polymer, $T$ the absolute temperature, $k_B$ and $R$
the Boltzmann and the ideal-gas constants.
In the dilute limit, $Z$ can be expanded in powers of the concentration as 
\begin{equation}
Z = 1 + \sum_{n\ge 2} A_n (c\hat{R}_g^3)^{n-1},
\label{Pi-exp}
\end{equation}
where $\hat{R}_g$ the zero-density radius of gyration. 
The coefficients $A_n$ depend on $T$, $N$, and on chemical details.
However, in the good-solvent regime, they converge to universal constants
$A_n^*$ as $N\to \infty$. Moreover, the renormalization group predicts
that corrections should always scale as $N^{-\Delta}$, where $\Delta$ is 
a universal exponent whose best estimate is\cite{BN-97} 
$\Delta = 0.515\pm 0.007^{+0.010}_{-0.000}$. Therefore, for large $N$
we expect
\begin{equation} 
  A_n(T,N) = A_n^* + A_{1,n}(T) N^{-\Delta} + \cdots
\label{AN-largeN}
\end{equation}
While the constants $A_n^*$ are universal, the coefficients $A_{1,n}(T)$ 
are system specific and temperature dependent. However, the ratios 
\begin{equation}
b_n \equiv {A_{1,n}(T)\over A_{1,2}(T) } 
\end{equation} 
are also universal. Precise estimates of $A_n$ for $n\le 4$ and 
of $b_3$ have been obtained in Ref.~\CITE{CMP-vir}:
\begin{eqnarray}
A_2^* &=& 5.500 \pm 0.003, \label{A2star} \\ 
A_3^* &=& 9.80  \pm 0.02,  \label{A3star} \\ 
A_4^* &=& -9.0  \pm 0.5,   \label{A4star} \\
b_3 &=& 4.75 \pm 0.30.
\end{eqnarray}
Instead of $A_2$ it is customary to define the interpenetration 
ratio\cite{footPsi} $\Psi \equiv 2 (4\pi)^{-3/2} A_2$,
whose large-$N$ value in the good-solvent regime is \cite{CMP-vir}
$\Psi^* = 0.24693\pm 0.00013$ (Ref.~\CITE{CMP-vir}),
$\Psi^* = 0.24685\pm 0.00011$ (Ref.~\CITE{PV-07}).

In the TPM, the coefficients $A_n$ become functions of $z$, $A_n(z)$, 
such that $A_n(z=0) = 0$ ($z=0$ is the ideal-chain case) 
and $A_n(z=\infty) = A_n^*$. Since $z\sim N^{1/2}$, 
Eq.~(\ref{AN-largeN}) implies 
\be
A_n(z) = A_n^* + a_n z^{-2\Delta}
\label{Anz-exp}
\end{equation}
for large $z$, with $a_n/a_2 = b_n$.\cite{footnote-bn}

The small-$z$ behavior of the TPM functions can be determined by 
using perturbation theory.\cite{Yamakawa-71,dCJ-book}
We quote here the result for 
$A_2(z)$ and $A_3(z)$:\cite{MN-87,Nickel-91,Casassa-72}
\begin{eqnarray}
A_2(z) &=& {1\over2} (4\pi)^{3/2} z [1 - 4.779663 z + 25.58964 z^2 + 
   O(z^3)], \label{A2-zexp} \\
A_3(z) &=& {2\over35} \left({16\pi\over3}\right)^3 
   (208 \sqrt{2} - 108 \sqrt{3} - 103) z^3 + O(z^4) \approx 
    1100.7 z^3 .
\label{A3-zexp}
\end{eqnarray}
In the appendix we compute the leading contribution to $A_4(z)$
obtaining 
\be
A_4(z) = {131072\over 45045} 
    (14075 + 12624 \sqrt{2} - 18468 \sqrt{3}) \pi^{9/2} z^4 + O(z^5) 
    \approx
    -29883.1 z^4.
\label{A4-zexp}
\ee
Beside the osmotic pressure we consider three different quantities that 
characterize the polymer size: the radius of gyration $R_g$, 
the hydrodynamic radius $R_H$, and the end-to-end distance. In the DJ lattice 
model they are defined as follows:
\begin{eqnarray}
R^2_g &\equiv & {1\over 2(N+1)^2} \left\langle \sum_{\alpha\beta}
   ({\mathbf r}_{\alpha} - {\mathbf r}_\beta)^2\right\rangle,
\\
{1\over R_H} &\equiv & {1\over (N+1)^2} \left\langle
     \sum_{\alpha\beta:{\mathbf r}_{\alpha}\not={\mathbf r}_\beta}
   {1\over |{\mathbf r}_{\alpha} - {\mathbf r}_\beta|} \right\rangle,
\\
R^2_e &\equiv & \langle ({\mathbf r}_{0} - {\mathbf r}_N)^2\rangle\; .
\end{eqnarray}
We also define the ratios
\begin{eqnarray}
A_{ge} \equiv  {\hat{R}_g^2\over \hat{R}_e^2}\qquad && \qquad 
A_{gH} \equiv  {\hat{R}_g\over \hat{R}_H},
\label{defratios}
\end{eqnarray}
where a hat indicates a zero-density quantity, and consider 
the density expansions
\begin{eqnarray}
{R^2_g\over \hat{R}_g^2} &=& 1 + S_{1,g} (c\hat{R}_g^3) +
       S_{2,g} (c\hat{R}_g^3)^2 + \cdots
\nonumber 
\\
{R^2_e\over \hat{R}_e^2} &=& 1 + S_{1,e} (c\hat{R}_g^3) +
       S_{2,e} (c\hat{R}_g^3)^2 + \cdots
\nonumber 
\\
{\hat{R}_H\over R_H} &=& 1 + S_{1,H} (c\hat{R}_g^3) +
       S_{2,H} (c\hat{R}_g^3)^2 + \cdots
\end{eqnarray}
The ratios (\ref{defratios}) and the density coefficient $S_{n,\#}$
are system-dependent quantities. However, as $N\to \infty$ in the 
good-solvent limit, they approach universal quantities, which 
will be labelled as $A_{ge}^*$, $A_{gH}^*$, and $S_{n,\#}^*$. 
The limiting values of the ratios and of the density coefficients for $n=1,2$
have been determined in Ref.~\CITE{CMP-raggi}. 

In the TPM all previous quantities are functions 
of $z$ which converge to their good-solvent value for $z\to \infty$.
If $Q$ corresponds to $A_{ge}$  
or to a coefficient $S_n$ for the radius of gyration and the end-to-end 
distance, we can also determine the corrections for $z\to\infty$. 
Indeed, in this limit, we have
\be
Q = Q^* + a_Q z^{-2\Delta}.
\label{Q-largez}
\end{equation}
The ratio $a_{Q}/a_2$ [$a_2$ is defined in Eq.~(\ref{Anz-exp})] 
is universal;
estimates for $A_{ge}$, $S_{1,g}$, and $S_{1,e}$ 
are reported in Ref.~\CITE{CMP-raggi}.
For $z\to 0$, $S_{n,\#}(z=0) = 0$, while $A_{ge}$ and $A_{gH}$ converge
to the ideal-chain (random-walk) values
\begin{equation}
A_{ge}(z=0) = {1\over6} \qquad\qquad A_{gH}(z=0) = {8\over 3\sqrt{\pi}}.
\end{equation}
The leading corrections for $z\to0$ 
to all these quantities are reported in the Appendix.

Finally, we consider the swelling factors for the zero-density radii:
\begin{eqnarray}
&& \hat{R}_g^2 = {1\over6} N\ell^2 \alpha_g^2(z), \\
&& \hat{R}_e^2 = N\ell^2 \alpha_e^2(z), \\
&& \hat{R}_H = {1\over8} \left({3\pi\over2}\right)^{1/2}  
    \sqrt{N} \ell \alpha_H(z) \; .
\end{eqnarray}
The swelling factors are normalized so that $\alpha_\#(z=0) = 1$.
In the DJ model the metrical factor $\ell$ is equal to the lattice spacing.
For $z\to \infty$ they behave as 
\begin{equation}
\alpha_\# = \alpha_{0,\#} z^{2\nu-1} (1 + \alpha_{1,\#} z^{-2\Delta} + \cdots)
\label{alpha-GS}
\end{equation}
For the hydrodynamic radius one should additionally consider corrections
proportional\cite{DRSK-02} to $z^{4\nu-4}$.

\section{Crossover functions} \label{sec4}

\subsection{Monte Carlo results} \label{sec4.1}

The main purpose of the present paper is the determination of the crossover 
functions for the quantities defined in Sec.~\ref{sec3}. 
We consider five different values of $z$, which we denote by 
$z_1$, $\ldots$, $z_5$,
which belong to the crossover region between ideal and good-solvent behavior.
Explicitly we use $z_1 = 0.056215$, 
$z_2 = 0.148726$, $z_3 = 0.32165$, $z_4 = 0.728877$, and $z_5 = 2.50828$. 
They were chosen so that $A_2(z_n) \approx n$ (remember that 
$A_2(z)$ varies between 0 and 5.50). In order to compute the TPM value
for each $z_i$, we perform several simulations at values 
$(w_{ij}, N_{ij})$ such that $w_{ij} N_{ij}^{1/2} = (2 \pi/3)^{3/2} z_i$,
choosing $N_{ij}$ between 100 and 8000. Then, we fit each universal quantity 
$Q$ with the theoretically expected behavior:\cite{BD-79,BN-97,foot-TPM}
\begin{equation}
Q(w_{ij},N_{ij}) = Q^*(z_i) + N_{ij}^{-1/2} b_Q(z_i) + 
       N_{ij}^{-1} c_Q(z_i).
\label{fit-Q}
\end{equation}
The TPM result corresponds to the leading term $Q^*(z_i)$. In order to detect 
additional scaling corrections that are not taken into account by the 
fit ansatz (\ref{fit-Q}), we have repeated the fit several times, each time 
including only data satisfying $N\ge N_{\rm min}$.

\begin{figure}
\begin{center}
\includegraphics[angle=-90,width=16truecm]{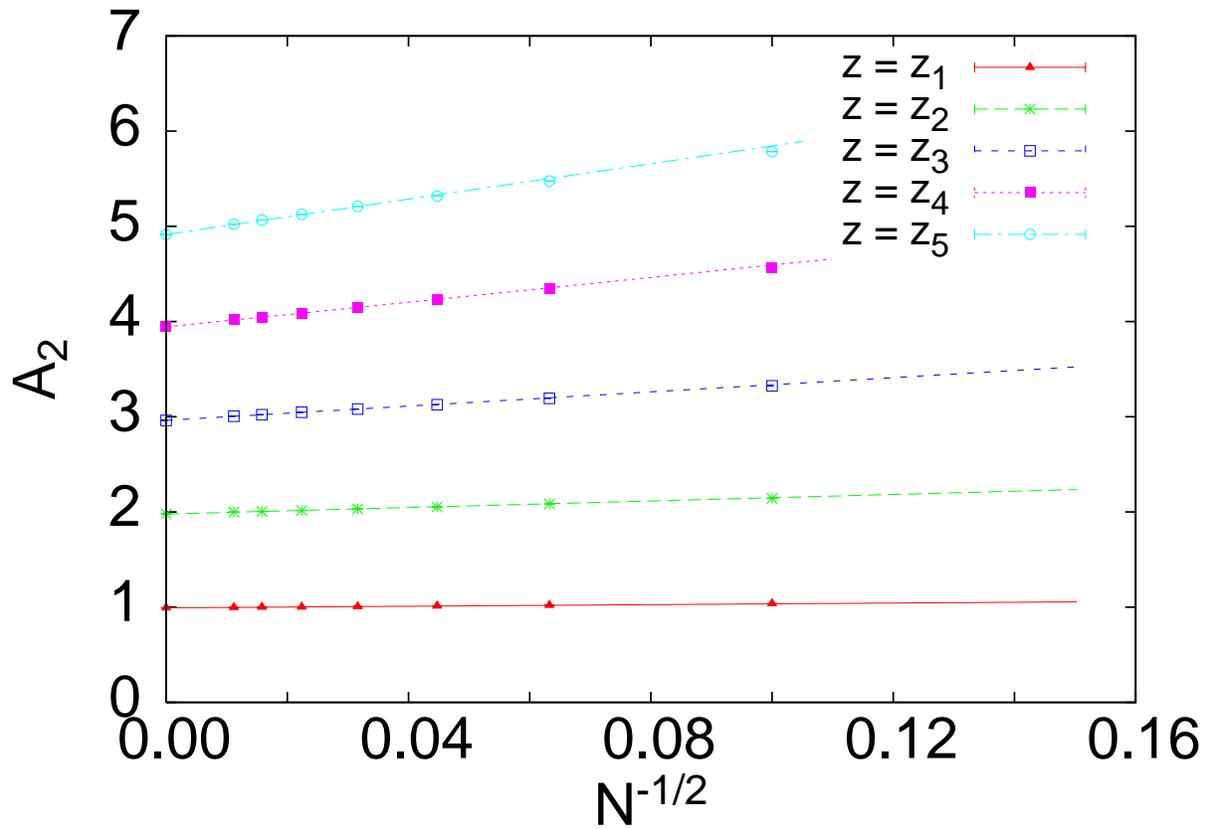}
\end{center}
\vspace{1cm}
\caption{The invariant ratio $A_2$ as a function of $N^{-1/2}$ 
for the five values of $z$ reported in the text.
We also report the linear extrapolation as determined by the fit of all
data.
}
\label{fig-A2}
\end{figure}

We illustrate the procedure by considering $A_2$. 
In Fig.~\ref{fig-A2} we plot $A_2(w,N)$ vs $N^{-1/2}$; 
for each $z$ we also plot the function $Q^* + N^{-1/2} b_Q$ 
obtained in the fit of all data ($N_{\rm min} = 100$)
to Eq.~(\ref{fit-Q}). The data points follow the expected behavior 
quite precisely, with very small $N^{-1}$ corrections. 
Note that an extrapolation is always needed except for very small values
of $z$. 
Estimates of $A_2(z)$ are reported in Table~\ref{tab-A2} for different values 
of $N_{\rm min}$. No systematic deviations are observed. We take the results
corresponding to $N_{\rm min} = 500$ as our final results. 
They are reported in Table~\ref{table-TPM}. We have applied the same analysis
to $A_3$, $A_4$, $A_{ge}$, $A_{gH}$, $S_{n,g}$, $S_{n,e}$, $S_{n,H}$
($n=1,2$). The results corresponding to $N_{\rm min} = 500$ are reported in
Table~\ref{table-TPM}.

\begin{figure}
\begin{center}
\includegraphics[angle=-90,width=10truecm]{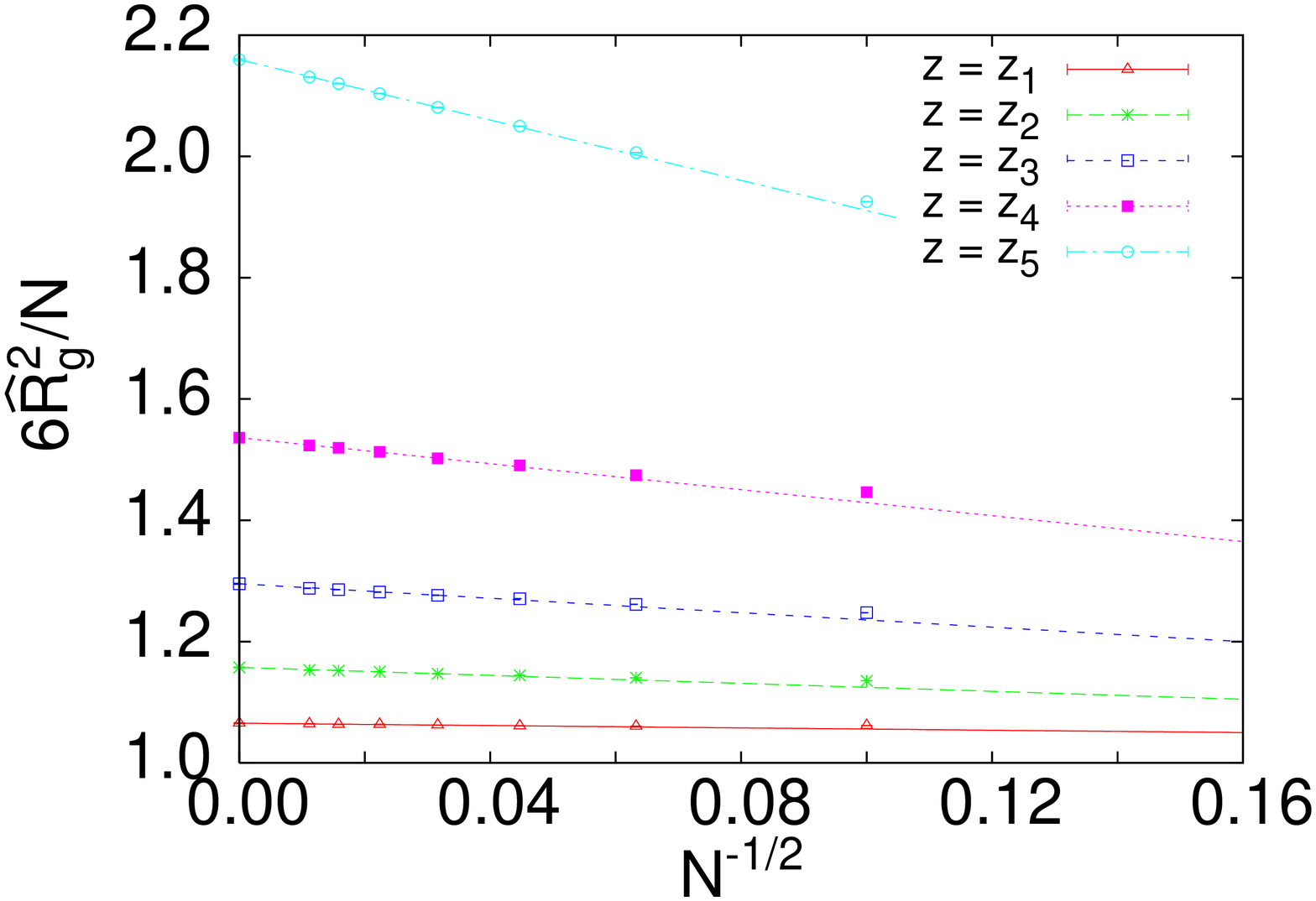}
\includegraphics[angle=-90,width=10truecm]{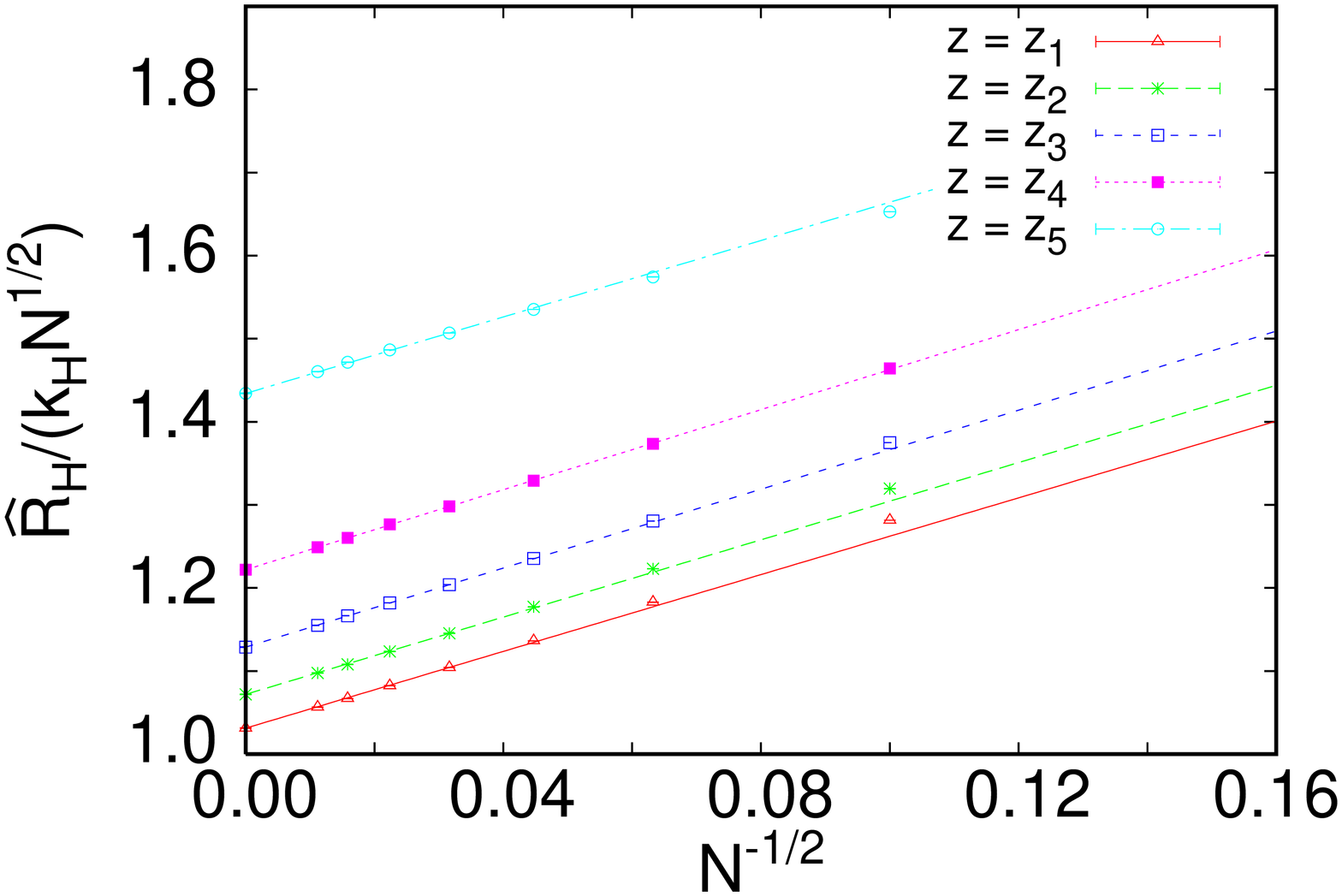}
\end{center}
\vspace{1cm}
\caption{The ratios $6\widehat{R}^2_g/N$ and $\widehat{R}_H/(k_H \sqrt{N})$ 
(they converge to $\alpha_g^2$ and $\alpha_H$, respectively) as a function 
of $N^{-1/2}$ for the five values of $z$ reported in the text.
We also report the linear extrapolation as determined by the fit of all
data.
}
\label{fig-swelling}
\end{figure}

Finally, we determine the swelling factors. We consider $6 \hat{R}_g^2/N$,
$\hat{R}_e^2/N$, and $k_H \sqrt{N}/\hat{R}_H$ 
($k_H = 2^{-7/2} (3 \pi)^{1/2}$). The corresponding quantities are reported in
Fig.~\ref{fig-swelling}. Their behavior with $N$ at fixed $z$ is perfectly
consistent with Eq.~(\ref{fit-Q}). Therefore, we have performed the same 
fits as before. The results are reported in Table~\ref{table-TPM}.

\subsection{Interpolation formulas} \label{sec4.2}

We now use the results of Sec.~\ref{sec4.1}, the good-solvent results 
of Refs.~\CITE{CMP-vir,CMP-raggi}, and the small-$z$ results mentioned 
in Sec.~\ref{sec3} and in the Appendix, to obtain interpolation formulas
that are valid for all values of $z$. We discuss in detail the virial
coefficients; all other quantities are analyzed analogously.

For the second virial coefficient we wish to find an interpolation that 
satisfies the following properties: (i) for $z\to \infty$ it 
must satisfy $A_2(z) \to A_2^* = 5.500$ [Eq.~(\ref{A2star})]; 
(ii) for $z\to 0$ it must behave as 
$4 \pi^{3/2} z (1 - 4.779663 z)$ [Eq.~(\ref{A2-zexp})];
(iii) the interpolating curve should assume the values determined numerically
and reported in Table~\ref{table-TPM}. Since the results for $A_2(z)$ 
indicate that this function is monotonic, we take 
an interpolating function of the form
\begin{equation}
A_2(z) = 4 \pi^{3/2} z 
  (1 + d_1 z + d_2 z^2 + d_3 z^3 + d_4 z^4)^{-1/4},
\label{A2-ansatz}
\end{equation}
where $d_i$ are constants to be determined. The constant $d_1$ can be fixed to
obtain the expansion (\ref{A2-zexp}) to order $z^2$: we obtain $d_1 = 19.1187$.
The constant $d_4$ can be fixed by requiring $A_2(z = \infty) = A_2^*$,
where $A_2^*$ is given in Eq.~(\ref{A2star}): this gives $d_4 = 268.96$. 
Then, we fit
\begin{equation}
 \left[ {4\pi^{3/2} z\over A_2(z)}\right]^4 - 1 - d_1 z - d_4 z^4 = 
 d_2 z^2 + d_3 z^3\; .
\end{equation}
Using the five data reported in Table~\ref{table-TPM} we obtain
$d_2 = 126.783$ and $d_3 = 331.99$. The interpolation formula is reported in 
Table~\ref{table-interp}. For $z\to \infty$, Eq.~(\ref{A2-ansatz}) gives 
\begin{equation}
A_2(z) = 5.500 - 1.6972/z + O(z^{-2})\; .
\label{A2interp-largez}
\end{equation}
This expression is compatible with Eq.~(\ref{Anz-exp}), taking into 
account that\cite{BN-97} $2 \Delta \approx 1.03$. It allows us to estimate
$a_2$: $a_2 \approx -1.7$. Of course, this is a very rough estimate.
A careful determination would require $A_2(z)$ for much larger values of $z$
and a careful analysis of the corrections to the 
behavior (\ref{Anz-exp}). Expression (\ref{A2interp-largez}) agrees with the 
field-theoretical result reported in Ref.~\CITE{Nickel-91}, which predicts
$a_2 \approx 5.50\times (-0.30) \approx - 1.65$.

We now compare the interpolation formula (\ref{A2-ansatz}) with
similar expressions that appear in the literature. We consider the expression
reported in Ref.~\CITE{Schaefer-99} (Sec. 15.5.2):
\begin{equation}
A_2(z) = {1\over2} (4\pi)^{3/2} 
  {0.182 \tilde{z} (1 + 2.15 \tilde{z} + 0.82 \tilde{z}^2)^{-0.236}
  \over (1 + 1.32 \tilde{z} + 0.378 \tilde{z}^2)^{0.264}},
\label{A2-sch}
\end{equation}
where\cite{foot-A2-Sch} $z = 0.182 \tilde{z}$. 
Equation (\ref{A2-sch}) has been 
obtained by using a sophisticated form of renormalized one-loop perturbation 
theory. By means of an extensive Monte Carlo simulation 
Ref.~\CITE{BN-97} obtained
\begin{equation}
A_2(z) = {1\over2} (4\pi)^{3/2} z 
 (1 + 14.339 z + 60.30 z^2 + 66.3 z^3)^{-1/3}.
\label{A2-Nick}
\end{equation}
Finally, we quote the field-theoretical 
expression of Ref.~\CITE{DF-84} for $\epsilon = 1$:
\begin{equation}
A_2(z) = {1\over2} (4\pi)^{3/2}
  \left\{ {\eta\over 8 (1 + \eta)} + 
    {1\over64} \left[ \left(4 \ln 2 + {7\over6}\right)
   \left({\eta \over 1 + \eta}\right)^2 + {21\eta\over 4 (1 + \eta)}
   \right]\right\},
\label{A2-Freed}
\end{equation}
with $\eta = 256 z/53$ (this relation is obtained by matching the small-$z$
behavior). Our result (\ref{A2-ansatz}) is essentially identical to 
Eq.~(\ref{A2-Nick}): differences are less than 0.3\%. It is also in very 
good agreement with the field-theoretical result (\ref{A2-sch}), differences 
being less than 1.5\%. Eq.~(\ref{A2-Freed}) is worse: the difference is 
of order 8\% for $z = 0.1$ and increases to 12\% for large values of $z$.

Let us now discuss $A_3$. We will use an interpolation formula analogous 
to (\ref{A2-ansatz}), setting
\be
A_3(z) = 1100.7 z^3 (1 + d_1 z + d_2 z^2 + d_3 z^3 + d_4 z^4)^{-3/4}.
\label{A3-ansatz}
\end{equation}
The prefactor has been fixed by using Eq.~(\ref{A3-zexp}). 
To determine the coefficients we use a strategy slightly different from that 
discussed for $A_2$, since we only know the leading small-$z$ 
behavior of $A_3$ and thus we cannot fix $d_1$ by using perturbation theory. 
Instead, we make use of the results of 
Ref.~\CITE{CMP-vir} to obtain the large-$z$ behavior 
of $A_3(z)$. Since\cite{CMP-vir} $b_3 = a_3/a_2 = 4.75$ 
[see Eq.~(\ref{Anz-exp})], Eqs.~(\ref{A2interp-largez}) and 
(\ref{A3star}) give $A_3(z) \approx 9.80-8.062/z$ 
(again we use the approximation
$2\Delta \approx 1$). If we require Eq.~(\ref{A2-ansatz})
to reproduce this expansion, we obtain
$d_3 = 594.386$ and $d_4 = 541.906$. Finally, we fit the results of 
Table~\ref{table-TPM} to determine $d_1$ and $d_2$. The resulting
expression is reported in Table~\ref{table-interp}. 

\begin{figure}
\begin{center}
\includegraphics[angle=-90,width=16truecm]{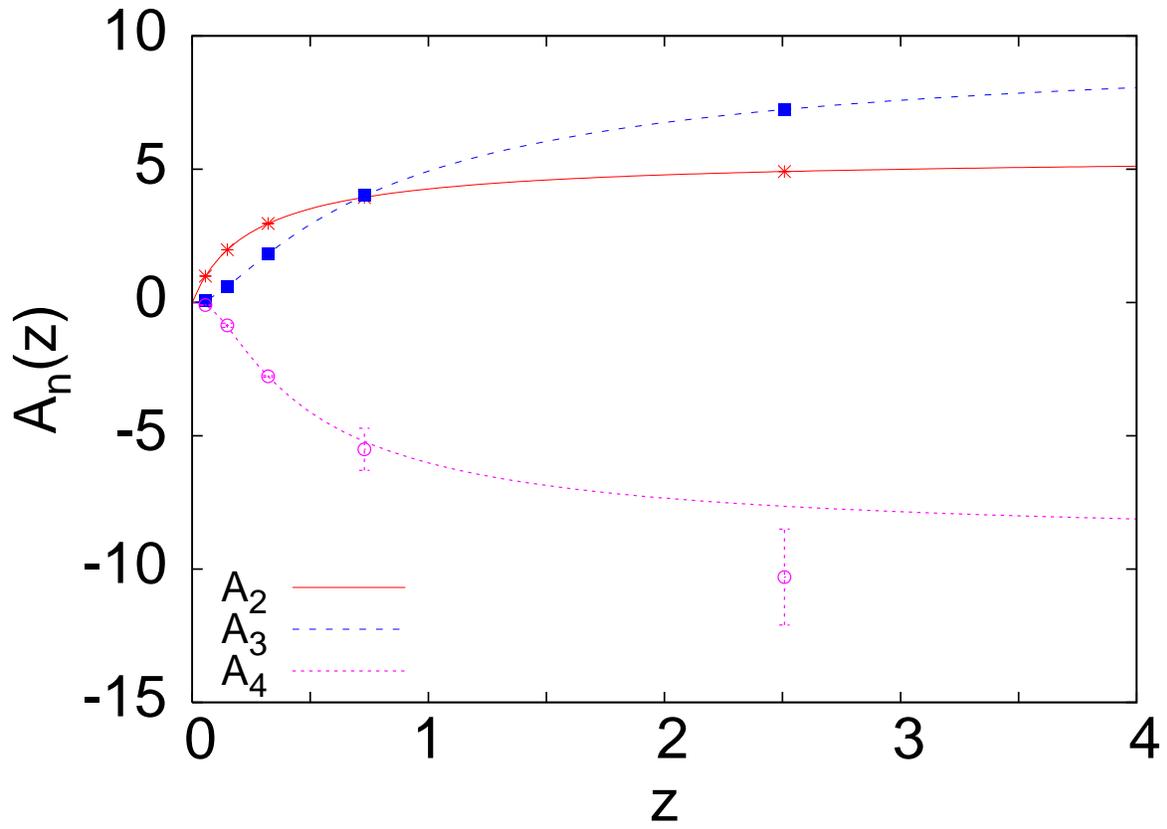}
\end{center}
\vspace{1cm}
\caption{The TPM functions for the virial coefficients.
We also show the Monte Carlo results reported in 
Table~\ref{table-TPM}. }
\label{fig-crossover-An}
\end{figure}

Finally, we consider $A_4$. In this case we do not know $a_4/a_2$ and thus
we use an interpolation formula with only three parameters:
\be
A_4(z) = - 29883.1 z^4 (1 + d_1 z + d_2 z^2 + d_3 z^3)^{-4/3}.
\end{equation}
The prefactor has been fixed by using Eq.~(\ref{A4-zexp}). The constant
$d_3$ is fixed by using Eq.~(\ref{A4star}), $d_1$ and $d_2$ by fitting the 
numerical results of Table~\ref{table-TPM}. The final expression is 
reported in Table~\ref{table-interp}. In Fig.~\ref{fig-crossover-An}
we report the crossover functions for $A_2$, $A_3$, $A_4$. They are monotonic
and approach the good-solvent value for $z \gtrsim 5$. 

\begin{figure}
\begin{center}
\includegraphics[angle=-90,width=9truecm]{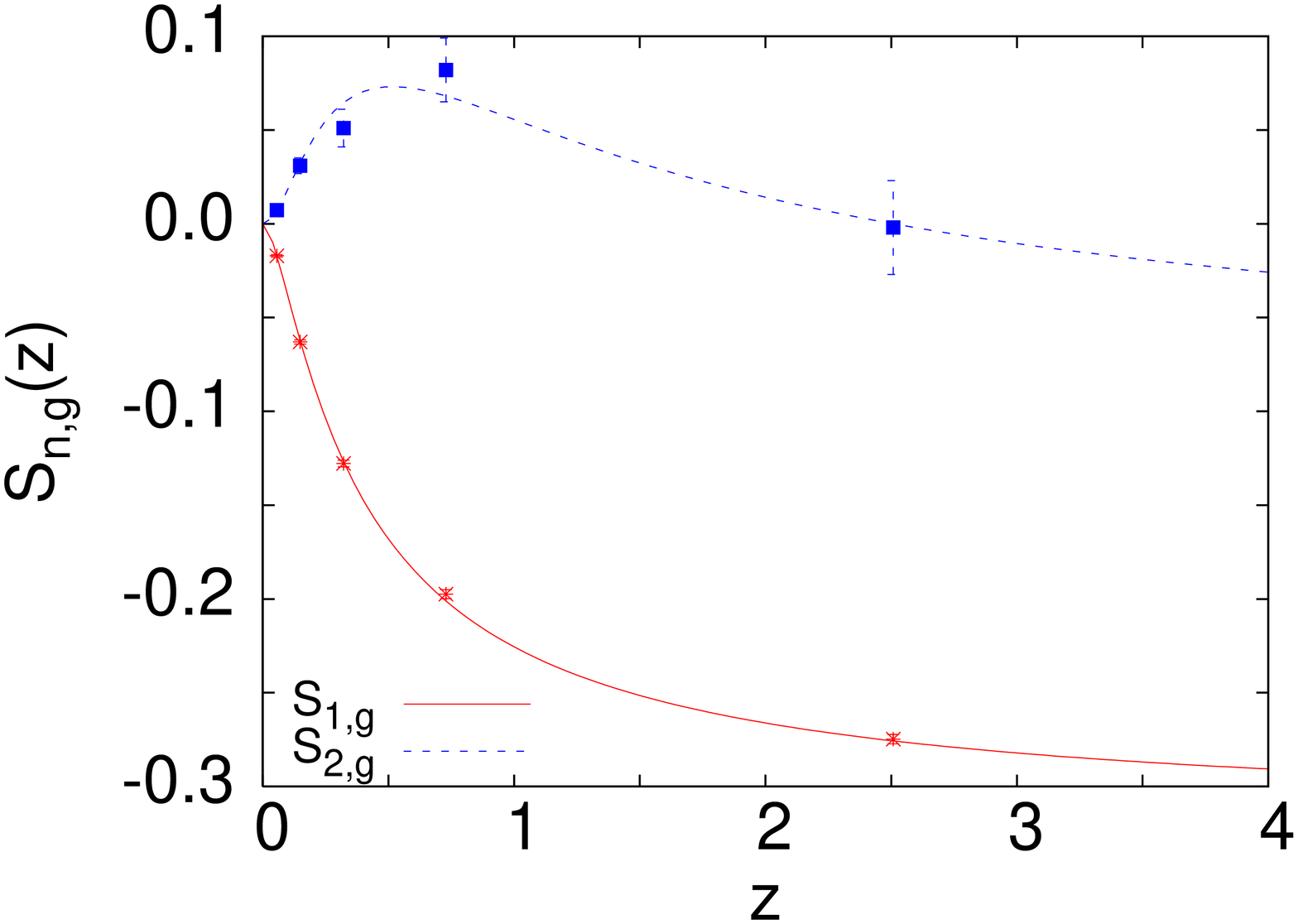}
\includegraphics[angle=-90,width=9truecm]{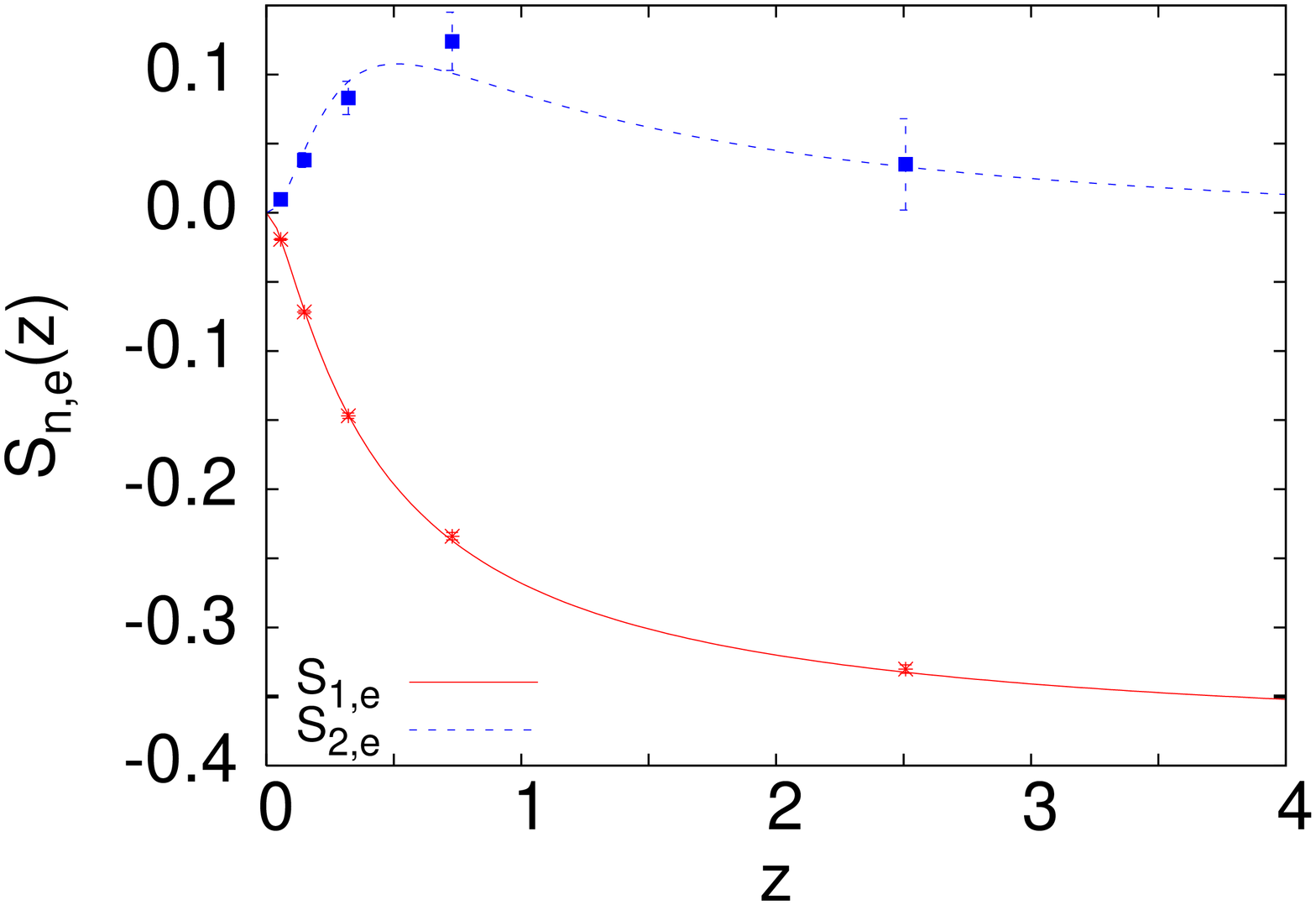}
\includegraphics[angle=-90,width=9truecm]{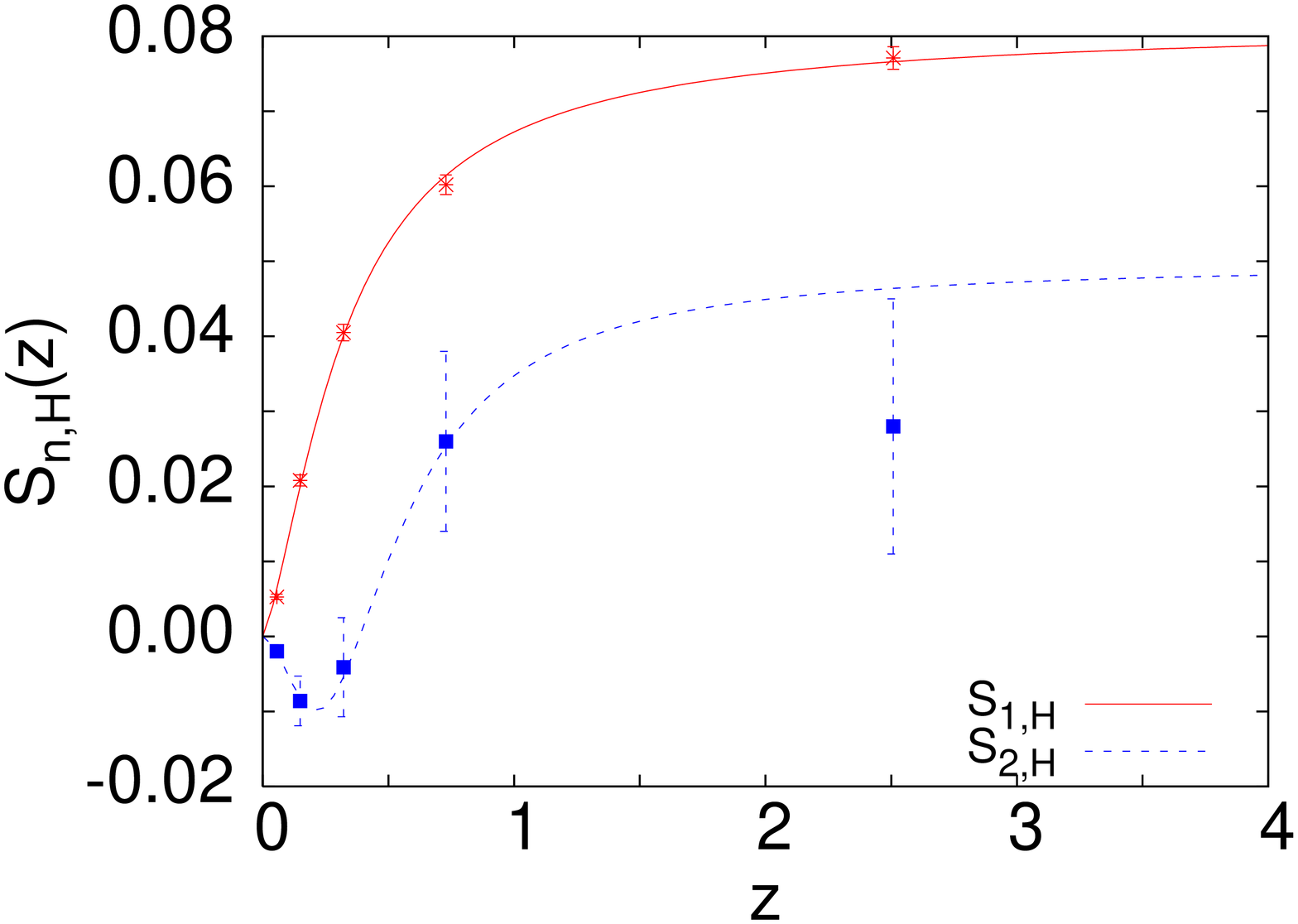}
\end{center}
\vspace{1cm}
\caption{The TPM functions for the density coefficients $S_{n,g}$, 
$S_{n,e}$, and $S_{n,H}$. 
We also report the Monte Carlo results reported in 
Table~\ref{table-TPM}. }
\label{fig-crossover-Sn}
\end{figure}

Similar analyses are performed for the density corrections to the 
radii. In this case, however, the crossover functions are not monotonic.
For the second density correction, this is evident from the numerical data.
For instance, $S_{2,g}$ and $S_{2,e}$ are first positive and increasing,
in agreement with Eqs.~(\ref{S2g-z}), (\ref{S2e-z}), reach a 
maximum for $0.5 \lesssim z \lesssim 1$, and then decrease,
converging to the good-solvent value which is negative. 
The density coefficient $S_{2,H}$ behaves in the opposite way, but note 
that, because of its definition, $S_{2,H}$ is equivalent in some sense to 
$-S_{2,e}$ and $-S_{2,g}$.
For the first density correction 
the nonmonotonicity can be inferred by using the results of 
Ref.~\CITE{CMP-raggi}. If 
$S_{1,\#} = S_{1,\#}^* (1 + \lambda_\# z^{-2\Delta})$, we have 
$\lambda_\# A_2^*/a_2 \approx - 0.050$ for both 
$S_{1,e}$ and $S_{1,g}$ [$a_2$ is defined in 
Eq.~(\ref{Anz-exp})]. Using $a_2 \approx - 1.697$ 
[Eq.~(\ref{A2interp-largez})], we obtain:
\be
S_{1,g} \approx -0.3152 - 0.0049/z^{2\Delta}, \qquad
S_{1,e} \approx -0.3853 - 0.0059/z^{2\Delta}.
\ee
Thus, the first density correction vanishes for $z = 0$, 
then decreases, becomes smaller than the good-solvent value, and 
eventually converges to it from below. 
This effect is however numerically very small
and thus the nonmonotonic approach is in practice irrelevant.
The nonmonotonic behavior requires interpolation formulas
slightly different from those used for the virial coefficients. 
For instance, expressions like (\ref{A2-ansatz}) cannot change sign 
and thus are unsuitable for $S_{2,\#}$. 
Our interpolations are reported in Table~\ref{table-interp} and 
plotted in Fig.~\ref{fig-crossover-Sn}. Note that the interpolations of 
$S_{2,\#}$
are not very precise in the region $0.5\lesssim z \lesssim 1$, 
since here the functions change their behavior and we do not have 
enough data points to identify precisely where the functions 
reach their maximum.

\begin{figure}
\begin{center}
\includegraphics[angle=-90,width=16truecm]{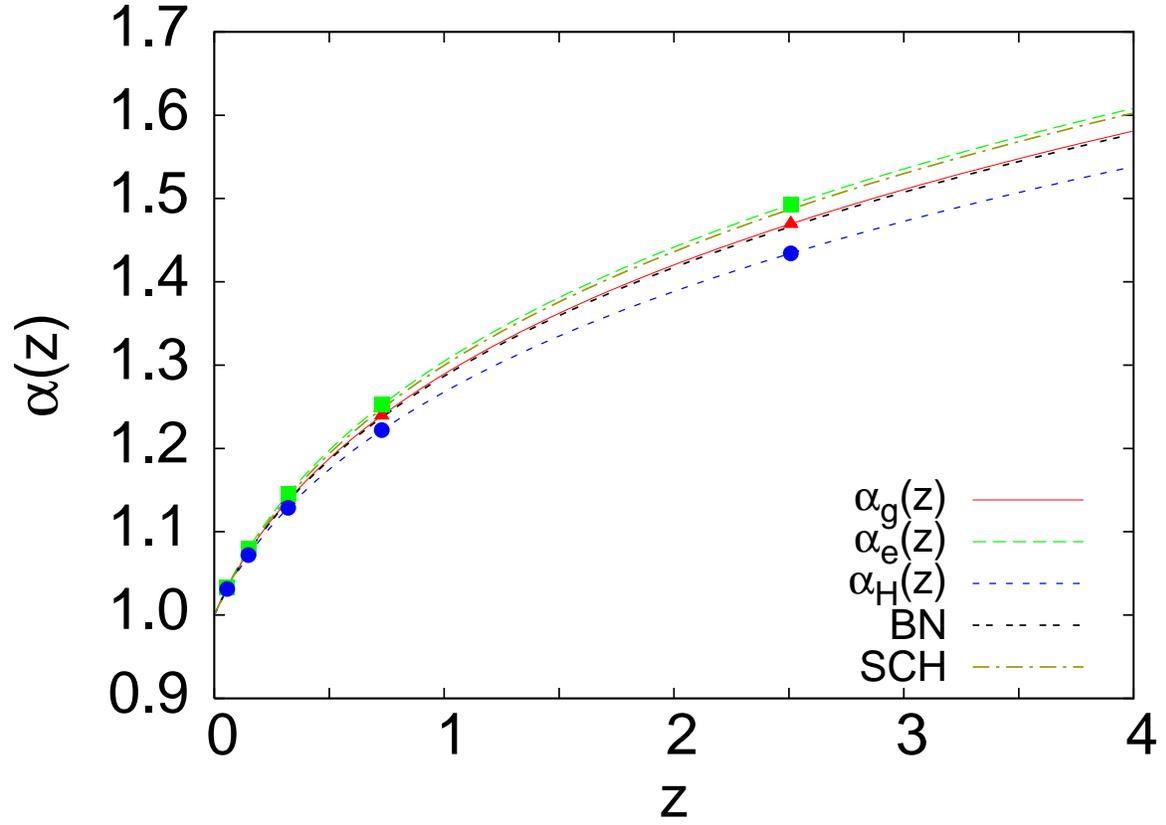}
\end{center}
\vspace{1cm}
\caption{The TPM functions for the swelling factors.
Together with our predictions we also report the results 
of Ref.~\protect\CITE{Schaefer-99} (SCH) and of 
Ref.~\protect\CITE{BN-97} (BN) for $\alpha_g(z)$.
Squares, triangles, and circles 
correspond respectively to the Monte Carlo results for 
$\alpha_e$, $\alpha_g$, and $\alpha_H$ (see
Table~\ref{table-TPM}). }
\label{fig-crossover-swelling}
\end{figure}

Finally, let us consider the swelling factors. 
Because of Eq.~(\ref{alpha-GS}), we use an interpolation of the form 
\begin{equation}
\alpha_\# = (1 + b_1 z + b_2 z^2 + b_3 z^3)^{(2\nu - 1)/3},
\end{equation}
taking\cite{BN-97} $\nu = 0.58758$. The coefficient $b_1$ 
is fixed by requiring $\alpha_\#$ to reproduce the small-$z$ 
behavior reported in the appendix, while $b_2$ and $b_3$ 
are obtained by interpolating the numerical data. 
The results are 
reported in Table~\ref{table-interp} and shown in 
Fig.~\ref{fig-crossover-swelling}. Note that $\alpha_e$,
$\alpha_g$, $\alpha_H$ behave in a very similar way, differences being tiny.
For $\alpha_g$ we also
show the prediction of Ref.~\CITE{BN-97}, 
$\alpha_g = (1 + 7.286 z + 9.51 z^2)^{0.087583}$,
and that of Ref.~\CITE{Schaefer-99}, 
$\alpha_g = (1 + 1.32 \tilde{z} + 0.378 \tilde{z}^2)^{0.088}$,
where\cite{foot-A2-Sch} $z = 0.182 \tilde{z}$.
The result of Ref.~\CITE{BN-97} is perfectly consistent with ours.
The field-theoretical result [note that in Fig.~\ref{fig-crossover-swelling}
it can hardly be distinguished from our result for $\alpha_e(z)$] 
is slightly larger (1\% at $z = 5$): differences are mainly related to the 
different choice of the exponent $\nu$. Finally, for the end-to-end distance
we mention the result of Ref.~\CITE{BN-97}:
$\alpha_e = (1 + 7.6118 z + 12.05135 z^2)^{0.087583}$.
Again, this expression is in perfect agreement with ours.

\subsection{The osmotic pressure} \label{sec4.3}

\begin{figure}
\begin{center}
\includegraphics[angle=-90,width=16truecm]{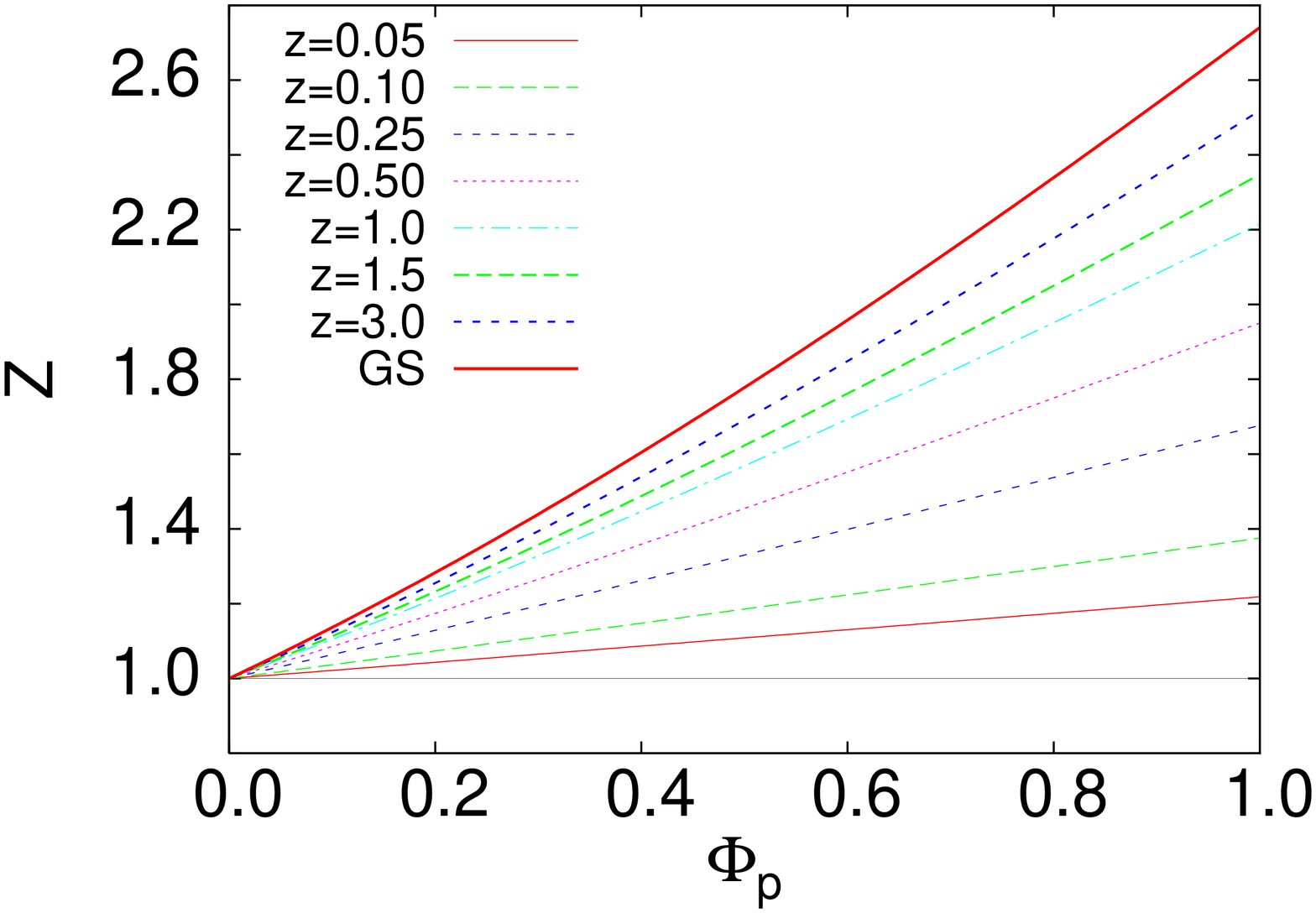}
\end{center}
\vspace{1cm}
\caption{The compressibility factor vs $\Phi_p$ for several 
values of $z$ in the dilute region.}
\label{fig-Z}
\end{figure}

Knowledge of the crossover functions for the lowest virial 
coefficients provides the osmotic pressure in the dilute regime
in which $\Phi_p\lesssim 1$, where $\Phi_p$ is the polymer 
packing fraction,
\begin{equation}
\Phi_p \equiv {4 \pi \hat{R}^3_g\over 3} c = 
              {4 \pi \hat{R}^3_g\over 3} {N_A\over M} \rho,
\end{equation}
$N_A$ the Avogadro number, $M$ the molar mass of the polymer,
$c$ and $\rho$ the number density and the weight  concentration, respectively.
In Fig.~\ref{fig-Z} we report the compressibility factor $Z$ defined in 
Eq.~(\ref{Pi-exp}) for several values of $z$ for $\Phi_p \lesssim 1$. 
In this range of concentrations the virial expansion converges quite well
\cite{CMP-vir} and thus our interpolations provide accurate 
estimates of $Z$ as a function of $z$ and $\Phi_p$.

In Ref.~\CITE{CMP-vir}
it was shown that a resummation of the virial expansion by using the known 
large-$\Phi_p$ behavior provides a reasonably accurate expression 
for $Z$ valid in the whole semidilute region. Here we apply the same
method to the determination of the leading $z^{-2\Delta}$ correction to the 
good-solvent value.  Since $A_n = A^*_n + a_n z^{-2\Delta}$ for large $z$,
we can write
\begin{equation}
Z(z,N,c) \approx Z^*(N,c) + z^{-2\Delta}  Z_1(N,c)\; .
\end{equation}
The functions $Z^*(N,c)$ and $Z_1(N,c)$ depend on $c$ and $N$. However,
for $N\to \infty$, the renormalization group predicts that they become 
universal functions of the packing fraction $\Phi_p$, so that 
\begin{equation}
Z(z,N,c) \approx Z^*(\Phi_p) + z^{-2\Delta}  Z_1(\Phi_p).
\label{Z-incr}
\end{equation}
The good-solvent function $Z^*(\Phi_p)$ is reported in Ref.~\CITE{CMP-vir}.
We will now determine the function $Z_1(\Phi_p)$. For this purpose we 
determine its large-$\Phi_p$ behavior. We expect 
\begin{equation}
   Z_1(\Phi_p) \sim \Phi_p^\alpha,\qquad\qquad \Phi_p\to\infty.
\label{Z1-largePhi}
\end{equation}
To fix $\alpha$, we note that $\Pi$ is a function of the monomer concentration
$c_m \equiv c N$ but not of the degree of polymerization $N$ 
for $c\to \infty$ (and therefore also $c_m\to\infty$).
Hence, $c z^{-2\Delta} \Phi_p^\alpha$ should be independent of $N$, 
once $c$ has been replaced by $c_m$. This condition implies 
\begin{equation}
\alpha = {\Delta + 1\over 3 \nu - 1}.
\label{exp-Z1-largePhi}
\end{equation}
In order to understand the region in the $(c,z)$ plane in which 
expansion (\ref{Z-incr}) is valid, we must discuss the 
expected scaling behavior in the large-concentration limit for generic 
values of $z$. In the TPM the osmotic
pressure satisfies the general scaling behavior\cite{Schaefer-99}
\begin{equation}
{\Pi\over k_B T} = c {\cal P}(c \ell^3 N^{3/2},z).
\end{equation}
In the limit $z \to 0$ the Flory-Huggins theory applies: for large
values of $c$, $\Pi$ is
proportional to the square of the monomer concentration $c_m$
and to the interaction strength $w$, so that 
\begin{equation} 
{\Pi\over k_B T} \sim w c_m^2 \ell^3 \sim
               c z (c \ell^3 N^{3/2}).
\end{equation}
For large $c$, the dependence on $N$ should disappear at fixed $c_m$ and $w$,
so that the relevant scaling variable is 
$z/(c \ell^3 N^{3/2}) = w/(c_m \ell^3)$. Therefore, we obtain the scaling
behavior
\begin{equation}
{\Pi\over k_B T} = z c^2 \ell^3 N^{3/2}
      f_Z\left( {z\over c \ell^3 N^{3/2}} \right) , 
\qquad\qquad 
Z = z c \ell^3 N^{3/2}
      f_Z\left( {z\over c \ell^3 N^{3/2}} \right) .
\label{Z-semidilute}
\end{equation}
The function $f_Z(x)$ is finite for $x\to 0$, while for $x\to \infty$, 
consistency with (\ref{Z-incr}) implies
\begin{eqnarray}
&& f_Z(x) \approx  a_f x^{(3\nu - 2)/(3\nu-1)} (1 + 
    b_f x^{-\Delta/(3\nu-1)}). 
\end{eqnarray}
The value $f_Z(0)$ can be determined by noting that for $z\to 0$ we can write
\be
Z \approx  z c \ell^3 N^{3/2} f_Z(0) = 
   A_2 c \hat{R}_g^3 = 4 \pi^{3/2} z c (N/6)^{3/2} \ell^3.
\ee
This implies $f_Z(0)= (2\pi/3)^{3/2}/2$.

Eq.~(\ref{Z-semidilute}) indicates that, at fixed large $z$, the compressibility
factor shows two different behaviors. If $\Phi_p$ is large but still
$z \gg c \ell^3 N^{3/2}$, $Z$ increases following 
Eq.~(\ref{Z-incr}). If the concentration is further increased, the argument
of $f_Z(x)$ decreases and eventually 
$Z \approx z c \ell^3 N^{3/2} f_Z(0) = 3 \sqrt{\pi} z \alpha_g^{-3}\Phi_p$, 
i.e. $Z$ becomes linear in the concentration. It is clear that this second 
regime cannot be obtained from extrapolations of results in the dilute 
region. We will thus consider only concentrations such that 
$z \gg c \ell^3 N^{3/2}$, so that we can use Eq.~(\ref{Z-incr}).

To determine the osmotic pressure for densities in the semidilute 
regime satisfying $z \gg c \ell^3 N^{3/2}$
we expand $A_n(z) = A_n^* + A_{n,1}/z$ for $z\to \infty$. Thus, we obtain 
for $z\to \infty$
\begin{eqnarray}
Z &\approx& 1 + 1.31303 \Phi_p + 0.558533 \Phi_p^2  - 0.122455 \Phi^3 
\nonumber \\ 
   &+ &
    {1\over z} \left(
     -0.405182 \Phi_p - 0.459468 \Phi_p^2  + 0.0507105 \Phi_p^3 \right) ,
\end{eqnarray}
which is consistent with (\ref{Z-incr}) if we approximate $2\Delta \approx 1$.
In Ref.~\CITE{CMP-vir} we determined an interpolation formula for the 
leading term with the correct large-$\Phi_p$ behavior. Here we do the same 
for the correction term:
we determine an interpolation formula that has the asymptotic 
behavior (\ref{Z1-largePhi}) 
for $\Phi_p\to \infty$ with $\alpha$ given by Eq.~(\ref{exp-Z1-largePhi}),
and agrees with the previous expansion for $\Phi_p\to 0$. 
A simple expression satisfying these two properties is 
\be
 - 0.405182 \Phi_p 
   (1 + 2.30016 \Phi_p + 1.08734 \Phi^2_p )^{0.493}\; .
\ee
Combining this expression with that obtained in Ref.~\CITE{CMP-vir}
we obtain for the compressibility factor
\begin{eqnarray}
Z &=& \left(
{1 + 1.52605 \Phi_p + 0.795366 \Phi_p^2 \over 1 + 0.5245 \Phi_p}\right)^{1.311} 
\nonumber \\
&& -  {0.405182\over z^{2\Delta}} \Phi_p 
   (1 + 2.30016 \Phi_p + 1.08734 \Phi^2_p )^{0.493}.
\label{Zlargez_interp}
\end{eqnarray}
It is not possible to determine {\em a priori}, for each $z$, the density range
$\Phi_p\lesssim \Phi_{p,\rm max}(z)$ 
in which Eq.~(\ref{Zlargez_interp}) applies. For $\Phi_p = 1$ we can 
compare expression (\ref{Zlargez_interp}) with the virial expansion
(\ref{Pi-exp}) including the terms up to $n=4$.
The relative difference is less than 5\% (1\%, 0.1\% respectively)
for $z\gtrsim 1.7$ (4, 12 respectively). Thus, for $\Phi_p = 1$, 
Eq.~(\ref{Zlargez_interp}) is substantially correct for $z\gtrsim 4$ and 
reasonably predictive for $z \gtrsim 2$. Accepting an error of 5\%
(it makes little sense to require a smaller error since 
our interpolation formulas cannot in any case be more precise than
5-10\%; see the discussion reported in Ref.~\CITE{CMP-vir}),
we can set 
$\Phi_{p,\rm max}(z=2) = 1$. An estimate of $\Phi_{p,\rm max}(z)$ for 
larger values of $z$ can be obtained by noting that 
\be
   \Phi_{p,\rm max}(z) \sim z^{6\nu -2} \sim z^{2.53}.
\label{Phipmax}
\ee
Indeed, 
Eq.~(\ref{Zlargez_interp}) is valid as long as
$z \gg c \ell^3 N^{3/2} \sim
c \hat{R}_g^3 \alpha_g^{-3} \sim \Phi_p \alpha_g^{-3}$.
Hence, since $\alpha_g \sim z^{2\nu-1}$ for large $z$, we obtain
Eq.~(\ref{Phipmax}). Therefore, with errors at most of 5\% we expect 
the range to extend up to $\Phi_{p,\rm max}(z) \approx (z/2)^{2.53}$.

\begin{figure}
\begin{center}
\includegraphics[angle=-90,width=16truecm]{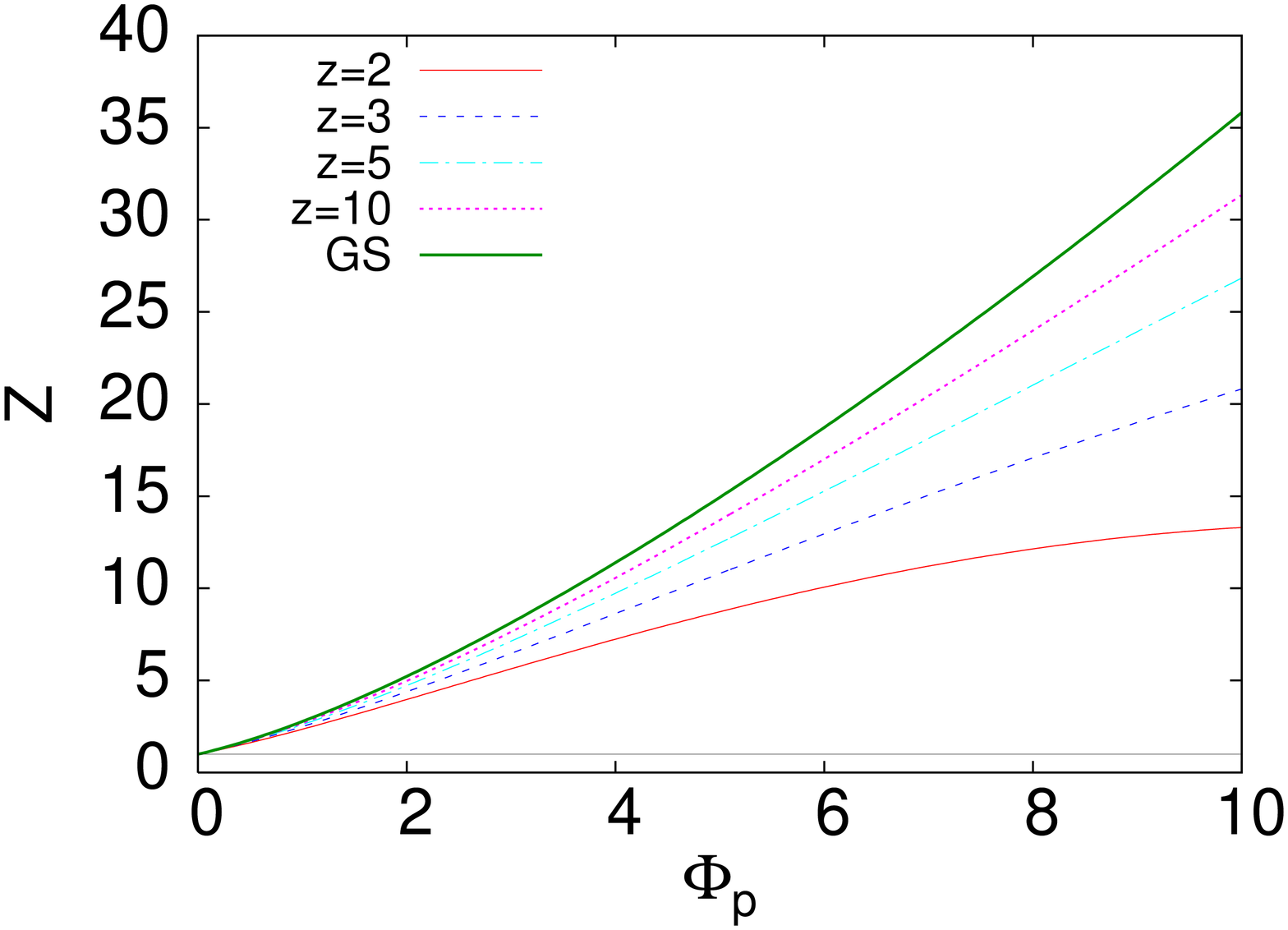}
\end{center}
\vspace{1cm}
\caption{The compressibility factor vs $\Phi_p$ for several 
values of $z$ in the semidilute region. }
\label{fig-Z-semidilute}
\end{figure}

The prediction (\ref{Zlargez_interp}) for $\Phi_p \le 10$ is reported in 
Fig.~\ref{fig-Z-semidilute} for several values of $z$. It is clear
that the results for $z = 2$ do not extend beyond $\Phi_p = 1$,
since, by increasing $\Phi_p$, $Z$ begins to bend in an unphysical way.
No such phenomenon is observed for $z \gtrsim 5$, which is therefore 
expected to be the range of $z$ in which (\ref{Zlargez_interp}) applies for 
$\Phi_p \lesssim 10$. This is in agreement with the estimate of 
$\Phi_{p,\rm max}(z)$ given above. 
Note that scaling corrections are quite large in the semidilute regime. 
For instance, consider $z = 10$. Since $A_2/A_2^* = \Psi/\Psi^* = 0.97$, 
in the dilute regime the solution is essentially in good-solvent conditions.
For $\Phi_p = 10$ we obtain $Z = 31.3$ to be compared with the 
good-solvent value $Z_{GS} = 35.8$: 
the pressure is lower by 14\%, a significant
deviation from the good-solvent value.

\subsection{Concentration dependence of the polymer size} \label{sec4.4}

\begin{figure}
\begin{center}
\includegraphics[angle=-90,width=16truecm]{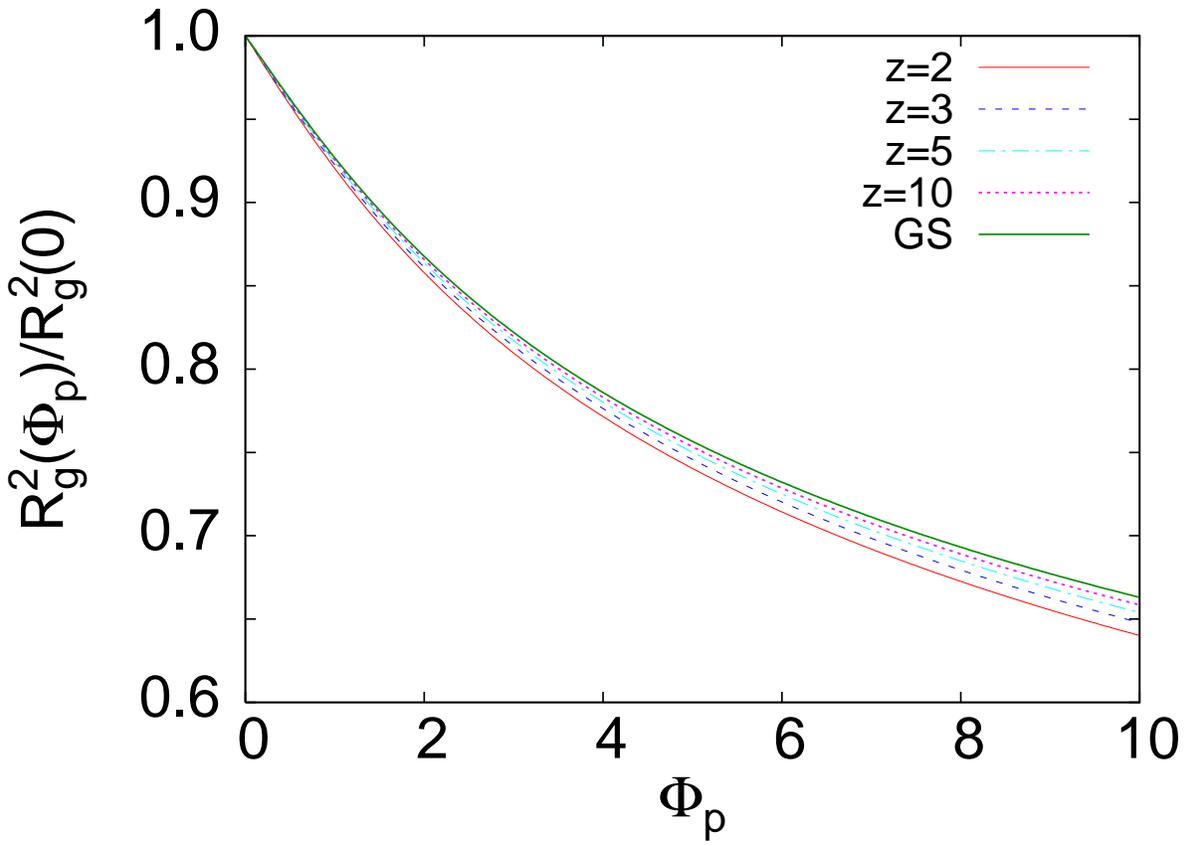}
\end{center}
\vspace{1cm}
\caption{Ratio $R^2_g/\hat{R}^2_g$ for the radius of gyration 
for several values of $z$ in the semidilute region. }
\label{fig-R-semidilute}
\end{figure}

The considerations we have presented for the osmotic pressure can be 
generalized to the radii. Deep in the semidilute region, polymers 
behave like ideal chains and therefore $R^2$ behaves as 
\begin{equation}
R^2 = N \ell^2 f_R (w/c_m) = 
    N\ell^2 f_R \left( {z\over c \ell^3 N^{3/2}} \right) ,
\label{R2-semidilute}
\end{equation}
with $f_R(0) \not=0$. As before, we focus on the deviations from 
the good-solvent
regime. If $R$ is the end-to-end distance or the radius of gyration,
corrections scale as $z^{-2\Delta}$ (this is not the case for the 
hydrodynamic radius which may show additional corrections proportional to 
\cite{DRSK-02,CMP-raggi} $z^{4\nu-4}$).
In order to obtain the behavior for large $c$ in the good-solvent regime
we write 
\begin{equation}
R^2 = a_R \hat{R}^2 \Phi_p^{\alpha_1} (1 + b_R z^{-2\Delta} \Phi_p^{\alpha_2}
     ),
\label{R2-extrap}
\end{equation}
and require this expression to be consistent with Eq.~(\ref{R2-semidilute}).
This allows us to identify $\alpha_1$ and $\alpha_2$:
\begin{equation} 
\alpha_1 = - {2\nu - 1\over 3\nu - 1} \qquad\qquad 
\alpha_2 = {\Delta \over 3\nu - 1}.
\end{equation}
Using these expressions, we can now extrapolate our virial results to the 
whole semidilute region, obtaining (the leading terms already appear 
in Ref.~\CITE{CMP-raggi})
\begin{eqnarray} 
{R^2_e\over \hat{R}^2_e} &=& 
  (1 + 0.801 \Phi_p + 0.37 \Phi^2_p )^{-0.115} - 
  {1\over z^{2\Delta}} 0.024 \Phi_p (1 + 0.76 \Phi_p)^{-0.554},
\\
{R^2_g\over \hat{R}^2_g} &=& 
  (1 + 0.655 \Phi_p + 0.28 \Phi_p^2)^{-0.115} - 
  {1\over z^{2\Delta}} 0.021 \Phi_p (1 + 1.46 \Phi_p)^{-0.554}.
\end{eqnarray}
As discussed before, these expressions are only valid for 
$z\ll c \ell^3 N^{3/2}$.
We do not present an extrapolation for $R_H$, because of the presence 
of two different corrections (one proportional to $z^{-2\Delta}$ and 
one proportional to $z^{4\nu-4}$, see Ref.~\CITE{DRSK-02}), 
which make extrapolations of the form (\ref{R2-extrap})
incorrect. The results for $R_g^2/\hat{R}^2_g$ are shown in 
Fig.~\ref{fig-R-semidilute}
for the same values of $z$ that occur in Fig.~\ref{fig-Z-semidilute}.
Note that for large $z$, our extrapolation predicts $R_g^2/\hat{R}^2_g$
to decrease when $z$ decreases. This is consistent with 
the nonmonotonic behavior of 
$S_{1,g}(z)$ we mentioned in Sec.~\ref{sec4.2} and 
with the numerical results of Ref.~\CITE{CMP-raggi}, which suggest
a negative scaling correction for $S_{2,g}(z)$:
$S_{2,g}(z) = - 0.087 - (0.003 \pm 0.005) z^{-2\Delta}$. 
Of course, this behavior should eventually change, since 
$R_g^2/\hat{R}^2_g$ should converge to 1 as $z\to 0$.

For the radius of gyration corrections are less evident than in the case of 
$Z$. For instance, for $\Phi_p = 10$ the relative difference between 
$R^2_g/\hat{R}^2_g$ for $z = 10$ and for $z=\infty$ (good-solvent value) is 
only of 0.7\%, indicating that the polymer size is less sensitive to the 
solvent quality far from the ideal-chain limit. 

\subsection{Comparison with previous renormalization-group results}
\label{sec4.5}

\begin{figure}
\includegraphics[angle=-90,width=16truecm]{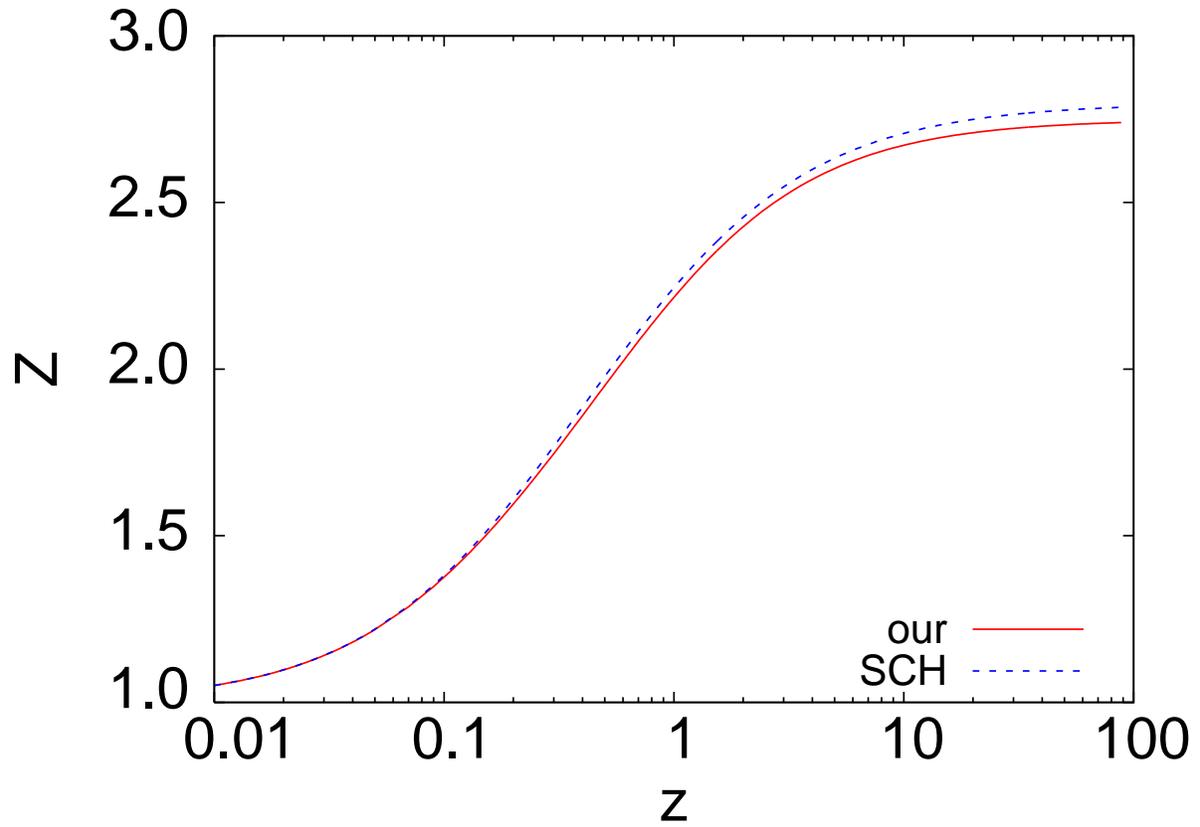}\vspace{1cm}
\begin{center}
\end{center}
\caption{The compressibility factor $Z$ at $\Phi_p = 1$ as a function of $z$.
We report our results ("our") and the field-theoretical ones 
reported in Ref.~\protect\CITE{Schaefer-99} ("SCH"). }
\label{fig-Z-phi1}
\end{figure}

\begin{figure}
\begin{center}
\includegraphics[angle=-90,width=16truecm]{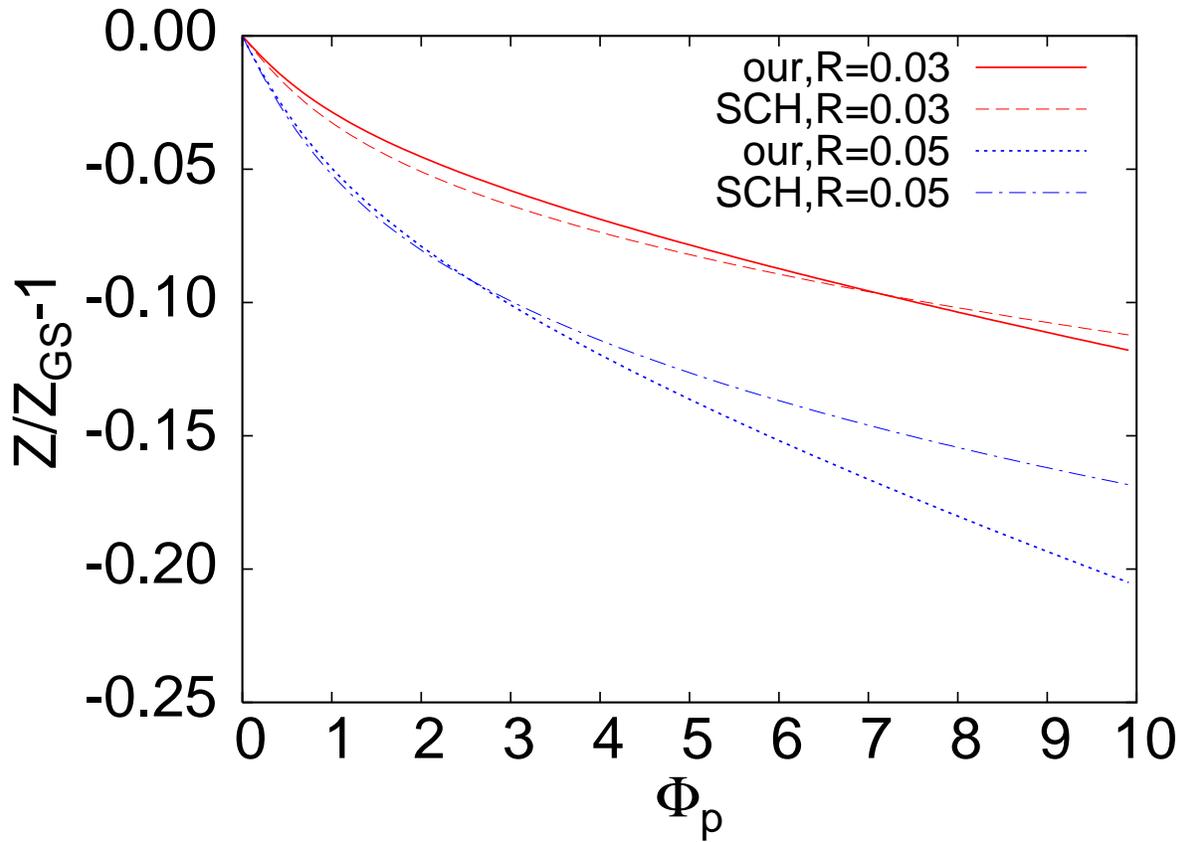}
\end{center}
\vspace{1cm}
\caption{Plot of $Z/Z_{GS}-1$ versus $\Phi_p$. $Z_{GS}$ is the compressibility
factor in the good-solvent regime, while $Z$ corresponds to 
solutions with two different values of $R \equiv 1 - {\Psi/\Psi^*}$.
We report the result (\ref{Zlargez_interp})  ("our")
and the field-theoretical one 
reported in Ref.~\protect\CITE{Schaefer-99} ("SCH"). }
\label{fig-Z-Phip-semidilute}
\end{figure}

In the previous sections we have obtained predictions for the osmotic 
pressure and the radii in the TPM. We wish now to compare 
these results with those obtained by using field theory. We compare mainly
with the results reported in Ref.~\CITE{Schaefer-99}, which have been
obtained by using renormalized one-loop perturbative expressions
with a careful choice of the renormalization constants. 
In Fig.~\ref{fig-Z-phi1} we report the compressibility factor $Z$ as a function
of $z$ for $\Phi_p=1$, i.e. for the largest value of the density at which 
the virial expansion is supposed to work. As we discussed in 
Ref.~\CITE{CMP-vir} this expression should be quite accurate, deviations 
being at most of 1-2\%. We also report the field-theoretical result, which is 
in perfect agreement with our estimate. For larger values of $\Phi_p$ we 
cannot use the virial expansion. Instead, we employ the approximate 
expression (\ref{Zlargez_interp}), which is valid only for large values of 
$z$. In Fig.~\ref{fig-Z-Phip-semidilute} we report $Z$ in the range 
$0\le \Phi_p \le 10$ for two values of $z$: $z = 9.90$ corresponding to 
$R\equiv 1 - \Psi/\Psi^* = 0.03$ and $z = 5.79$ corresponding to 
$R = 0.05$. We report our expression (\ref{Zlargez_interp}) and the 
field-theoretical prediction of Ref.~\CITE{Schaefer-99}. For $R = 0.03$,
our result is in very good agreement with the field-theoretical one.
On the other hand, for $R = 0.05$, we observe significant differences 
for $\Phi_p\gtrsim 5$. In any case, these discrepancies are within
the 5\% error we expect on our extrapolations. 
Indeed, for $\Phi_p = 5$ we predict $Z_{GS} = 15.0$ and 
$Z(z=5) = 12.9$ to be compared with the field-theory results
$Z_{GS} = 14.5$ and $Z(z=5) = 12.7$; for $\Phi_p = 10$ we 
obtain $Z_{GS} = 35.9$ and
$Z(z=5) = 28.5$ to be compared with 
$Z_{GS} = 34.2$ and $Z(z=5) = 28.4$. In all cases differences 
are less than 5\%.
Finally, note that both field theory and our results predict
$Z$ to be significantly lower than the good-solvent value $Z_{GS}$
for $\Phi_p \gtrsim 1$ as soon as $R$ is different from zero. 

A phenomenological expression for $Z$ as a function of $c$ 
and $R = 1 - \Psi/\Psi^*$ is also reported in Refs.~\CITE{LK-02,Lue-00}.
For $\Phi_p\lesssim 1$ there is good agreement \cite{footnoteLue} 
with our results
for all values of $R$, differences being less than 1\%. On the other 
hand, significant differences are observed for larger values of $\Phi_p$.
\cite{footnoteLue} For instance, for $R = 0.03$ and $\Phi_p = 10$ we predict
$Z/Z_{GS} - 1 = -0.118$, Ref.~\CITE{Schaefer-99} gives $-0.113$, in substantial
agreement with our result, while the expression reported in 
Ref.~\CITE{LK-02} gives $-0.060$, which differs by a factor of 2. 

\begin{figure}
\begin{center}
\includegraphics[angle=-90,width=16truecm]{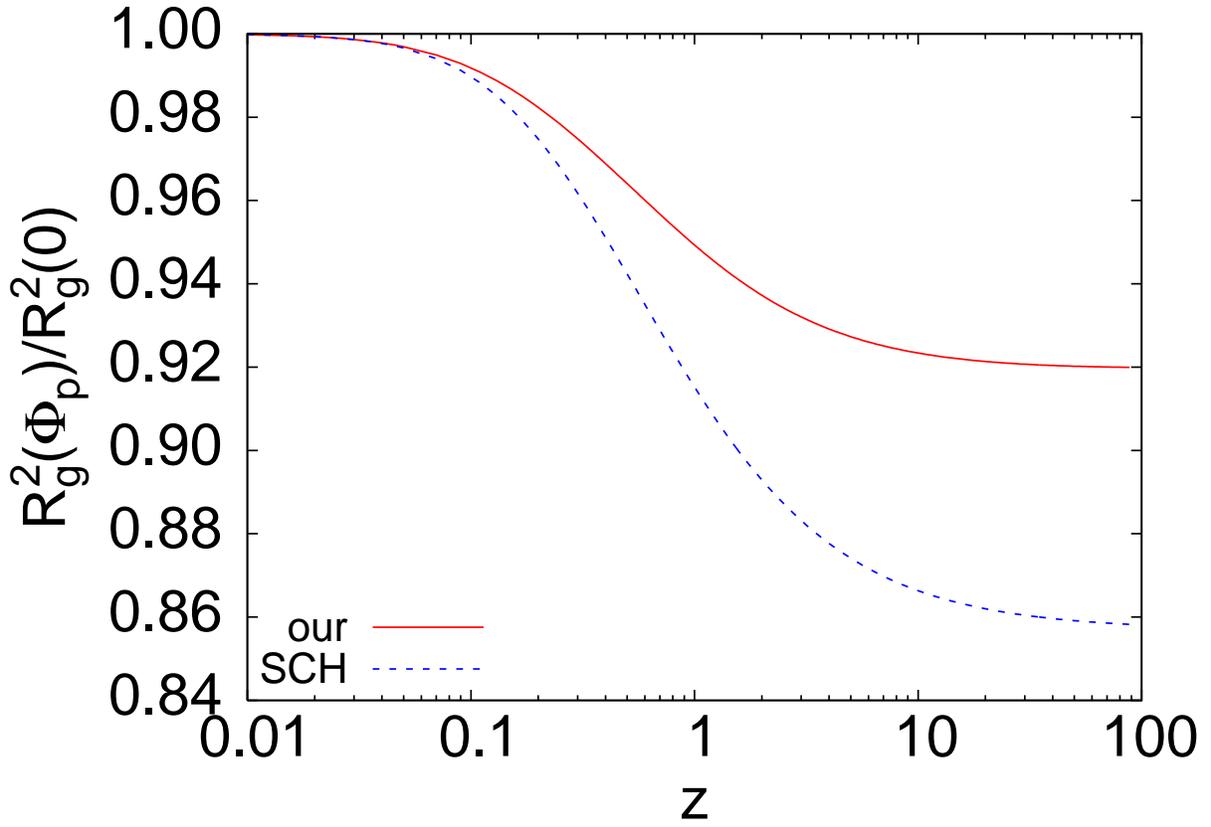}
\end{center}
\vspace{1cm}
\caption{The ratio $R^2_g/\hat{R}_g^2$ 
at $\Phi_p = 1$ as a function of $z$.
We report our results ("our") and the field-theoretical ones 
reported in Ref.~\protect\CITE{Schaefer-99} ("SCH"). }
\label{fig-Rg2-phi1}
\end{figure}

It is also interesting to compare the results for ${R}^2_g/\hat{R}^2_g$.
In Fig.~\ref{fig-Rg2-phi1} we report this ratio for $\Phi_p = 1$ as a function
of $z$, together with the one-loop prediction of Ref.~\CITE{Schaefer-99}.
In the good-solvent regime the difference is quite large. The poor
behavior of the field-theoretical expressions in the dilute regime 
can be explained by looking at the virial expansion of $R_g^2$ 
for $z\to\infty$. 
In the good-solvent regime, the Monte Carlo simulations of 
Ref.~\CITE{CMP-raggi} give $S_{1,g}^*\approx -0.315$, 
$S_{2,g}^*\approx -0.09$, while Ref.~\CITE{Schaefer-99}
predicts $S_{1,g}^*\approx -1.19$, $S_{2,g}^*\approx 5.74$.
One-loop perturbation theory renormalized as in Ref.~\CITE{Schaefer-99} 
seems to be unable to reproduce correctly the polymer size as a function of 
$\Phi_p$, at variance with what occurs for $Z$. This is not totally surprising
since the nonuniversal parameters that enter in the perturbative predictions
($c_0$ and $n_0$ in the notation of Ref.~\CITE{Schaefer-99}) were tuned 
to reproduce accurately the thermodynamic behavior.\cite{footnote_c0} 
For instance,
a different resummation of one-loop perturbation theory 
(see Sec.~6 of Ref.~\CITE{ON-83}) gives 
$S_{1,g}^*\approx -0.145$, $S_{2,g}^*\approx 0.0738$, 
which are much closer to the Monte Carlo results.

\section{Conclusions} \label{sec5}

In this paper we have determined the explicit TPM expressions for several
quantities which characterize polymer solutions in the dilute regime.
First, we have determined the universal constants $A_n(z)$ for $n=2,3,4$. 
This allows us to obtain precise predictions for the osmotic pressure
in terms of the polymer packing fraction $\Phi_p$ in the dilute regime
$\Phi_p \lesssim 1$. Then, we have computed the swelling factors
for three different radii that characterize the polymer size.
Finally, we have studied their concentration dependence.

The expressions we have determined in this paper can be used in two different
contexts. First, they provide expressions that may be used to fit 
the experimental data outside the universal (large-$N$) regime. 
Note that the range of values of $N$ in which a given system approximately 
behaves as predicted by the TPM expressions is nonuniversal, and 
thus in some cases the agreement is only at the level of the leading behavior,
while in some others it may cover a significant range of polymer lengths.
For instance, in some systems (e.g., in 
PMMA in chloroform or nitroethane at $20$ $^\circ$C) the interpenetration 
ratio approaches the universal value from above, i.e. 
$\Psi > \Psi^*$ for large $N$ (in the terminology of Ref.~\CITE{Schaefer-99},
these systems are "strong-coupling systems").
This type of behavior cannot be described 
by the TPM expressions we have derived here which predict that the approach
is always from below. In some other systems (``weak-coupling systems") 
instead, the TPM expressions
describe quite well the experimental behavior (for instance, 
polystyrene in cyclohexane or transdecaline, see chapter~15 in 
Ref.~\CITE{Schaefer-99}). Note also that $\Psi$ is generically a 
decreasing function of the temperature (at $T\to T_\theta$, $\Psi\to 0$)
and thus a strong-coupling system becomes a weak-coupling system as
$T$ is lowered and thus eventually one can use the TPM to interpret the 
experimental behavior.

\begin{figure}
\begin{center}
\begin{tabular}{cc}
\includegraphics[angle=-90,width=8truecm]{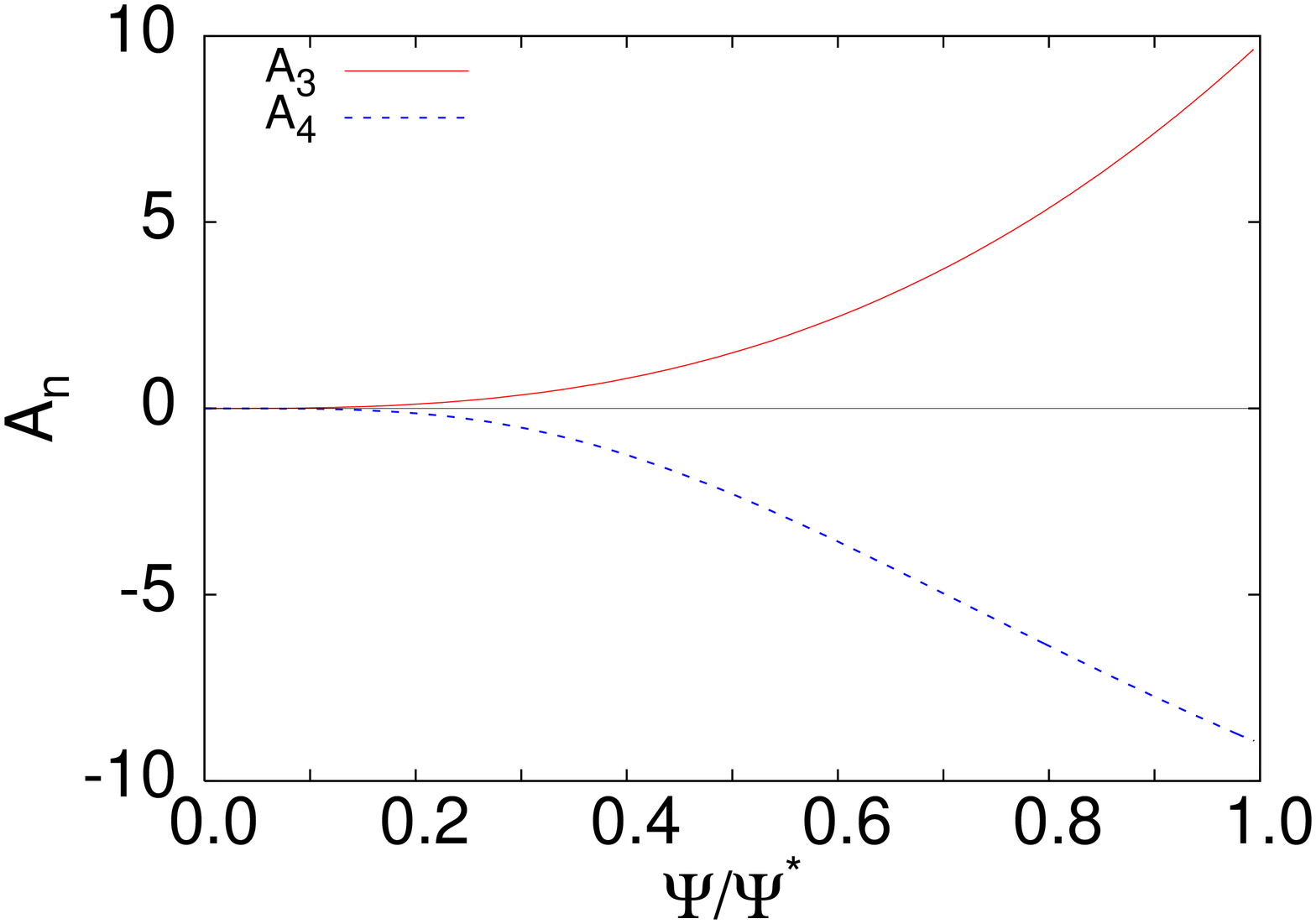} &
\includegraphics[angle=-90,width=8truecm]{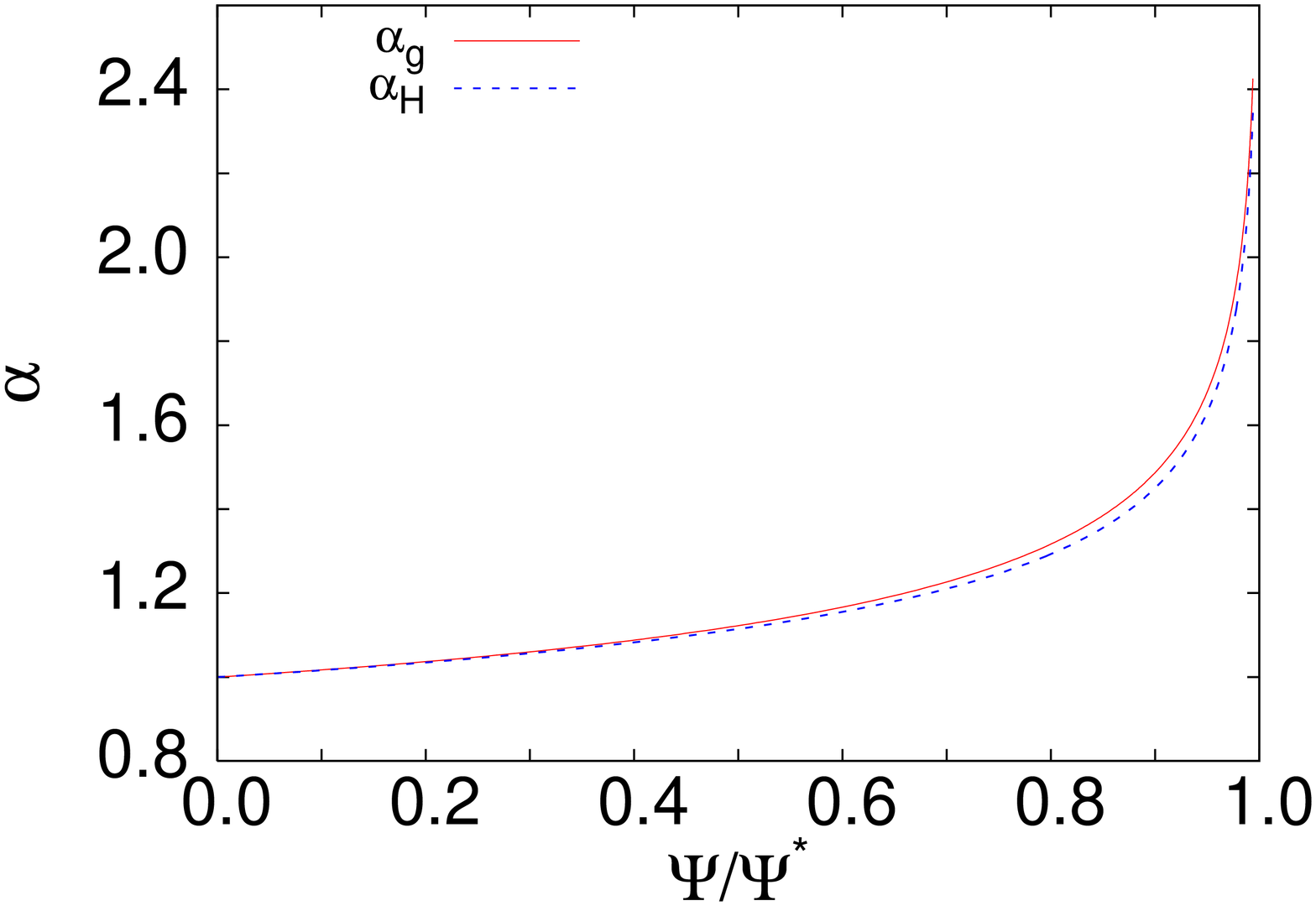} \\
\includegraphics[angle=-90,width=8truecm]{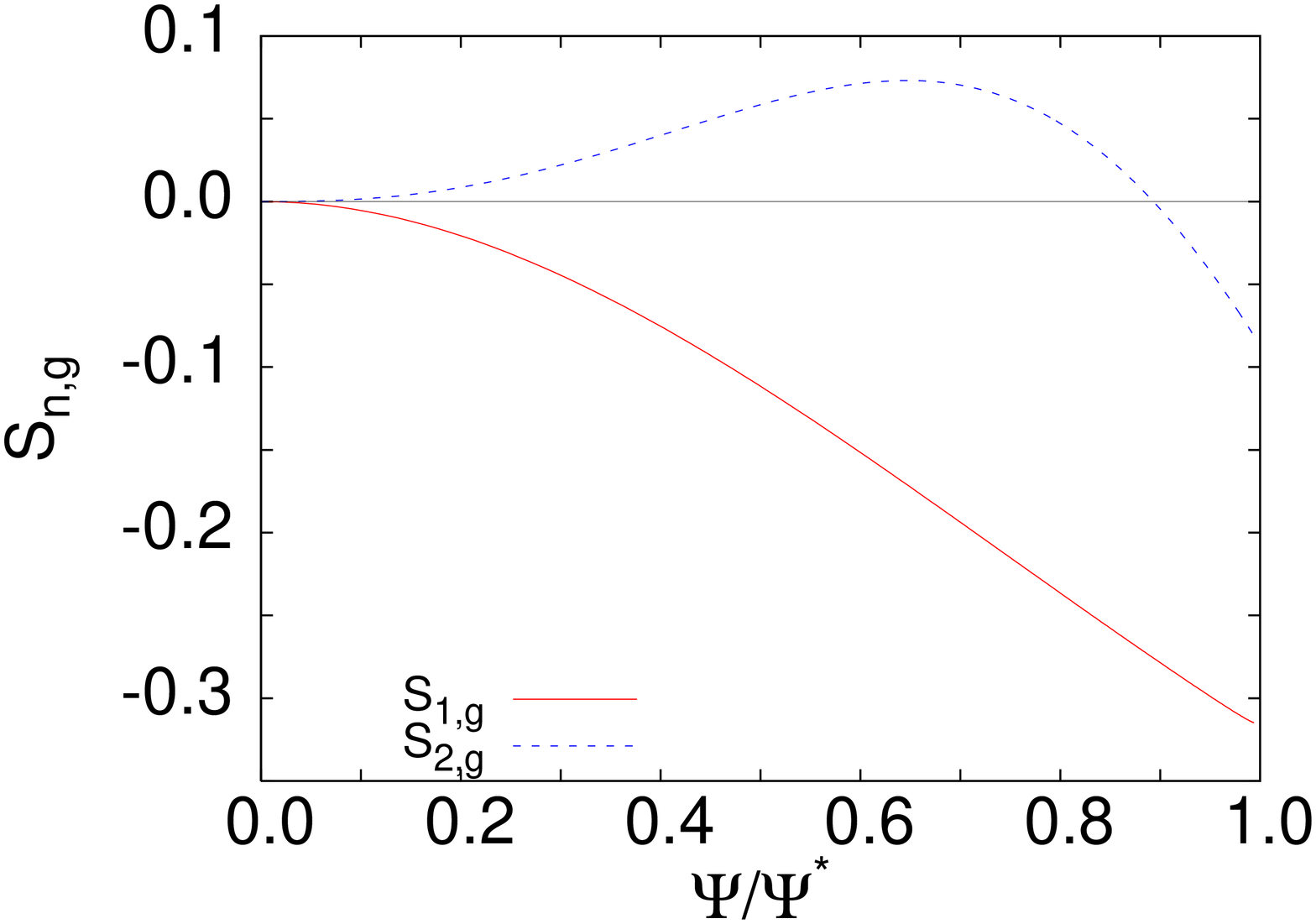} &
\includegraphics[angle=-90,width=8truecm]{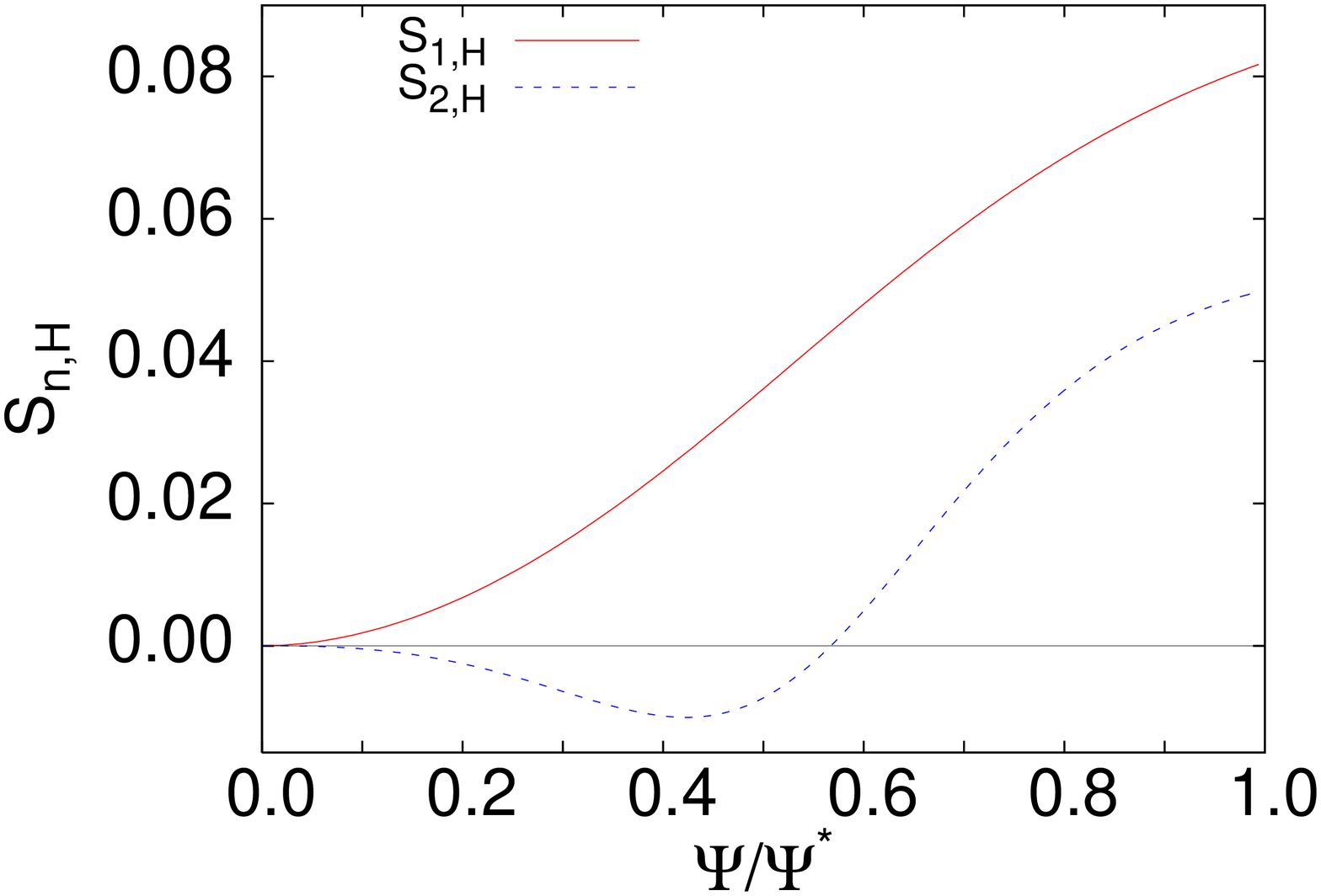} \\
\end{tabular}
\end{center}
\vspace{1cm}
\caption{The crossover functions vs $\Psi/\Psi^*$. We report:
(top left) $A_3$ and $A_4$; (top right) $\alpha_g$ and $\alpha_H$;
(bottom left) $S_{1,g}$ and $S_{2,g}$; (bottom right) $S_{1,H}$ and $S_{2,H}$.}
\label{fig-theta}
\end{figure}

The use of the TPM to describe the nonasymptotic behavior for finite values of
$N$ is mainly phenomenological and rigorously justified only in systems in 
which the persistence length is larger than the typical monomer size.
\cite{CCPRV-01} A second use of the TPM expressions is in the description
of the crossover to the $\theta$ point. As we have explained 
in the introduction, the crossover functions defined 
in Eq.~(\ref{scaling-gen}) can be computed in the TPM, by identifying 
$z$ with $(T-T_\theta) N^{1/2} (\ln N)^{-4/11}$ (modulo a normalization
multiplicative constant). However, this identification is valid only
close to the $\theta$ point, since it assumes $T-T_\theta\ll 1$.
To avoid this limitation, one can proceed as suggested in 
Refs.~\CITE{Nickel-91,PH-05},
i.e., one can parametrize the crossover in terms of a physical variable. For
instance, one can use the interpenetration ratio $\Psi$. Eq.~(\ref{scaling-gen})
can then be written as 
\begin{equation}
{\cal O}(T,N,c) = \alpha_1 {\cal O}_G(N,c) f_{\cal O} (\Psi,c\hat{R}_g^3).
\end{equation}
The {\em quality} of the solution is now characterized by $\Psi$
that varies between 0 (poor solvent) and $\Psi^*$ (good solvent).
In Fig.~\ref{fig-theta} we report the quantities computed above
as a function of $\Psi/\Psi^*$. Note that both $A_n$ and $S_{n,\#}$ 
are small up to $\Psi/\Psi^*\approx 0.3$ and also the swelling 
factors $\alpha$ do not change significantly in this range. 
This means that, for $\Psi \lesssim 0.3 \Psi^* \approx 0.08 $,
polymers behave approximately
as Gaussian coils. Of course, as $T\to T_\theta$ three-body
forces become increasingly important and thus tricritical corrections
should be included. In the opposite range $\Psi \gtrsim 0.08$ 
tricritical effects can be neglected (see the numerical data in 
Ref.~\CITE{PH-05}) and one can use the TPM expressions to describe the 
polymer behavior.

\vskip 1truecm

The authors thank Tom Kennedy for providing his efficient simulation code 
for lattice self-avoiding walks.

\appendix
\section{Perturbative calculations} \label{appendix}

In this appendix we report the one-loop TPM expressions for the 
quantities reported in this paper. We use the general results of 
Ref.~\CITE{ON-83}. For the osmotic pressure we start from the 
one-loop expression
\be
{\beta \Pi \over c} = 1 + {u\over2} c n^2 - {1\over 2c} 
 \int {d^3p\over (2\pi)^3} \ln [1 + 2 c u \Gamma^{(2)}(p)] + 
u \int {d^3p\over (2\pi)^3} {\Gamma^{(2)}(p)\over 1 + 2 c u \Gamma^{(2)}(p)},
\label{ON-int}
\ee
where $u$ is the coupling constant and $n$ is the polymer length. 
They are related to $z$ and $R_g$ by 
\be
z = (2\pi)^{-3/2} u n^{1/2}\qquad\qquad 
\hat{R}_g^2 = {n\over2} + O(u).
\ee
The function $\Gamma^{(2)}(p)$ is the Debye function:
\be
{1\over n^2} \Gamma^{(2)}(p) = {2\over p^2 n} - {4\over (p^2 n)^2} 
  (1 - e^{-p^2 n/2}).
\ee
Expanding in powers of $c$ we obtain $A_4$: 
\be
A_4(z) = {131072\over 45045} 
    (14075 + 12624 \sqrt{2} - 18468 \sqrt{3}) \pi^{9/2}    z^4 + O(z^5) = 
    -29883.1 z^4 + O(z^5).
\ee
We give also numerical values for the leading TPM contributions to the 
following virial coefficients:
\begin{eqnarray}
  A_5(z) &=& 932283 z^5, \\
  A_6(z) &=& -3.13006\cdot 10^7\, z^6, \\
  A_7(z) &=& 1.10136\cdot 10^9\, z^7, \\
  A_8(z) &=& -4.00539\cdot 10^{10}\, z^8, \\
  A_9(z) &=& 1.49307\cdot 10^{12}\, z^9, \\
  A_{10}(z) &=& -5.67361\cdot 10^{13}\, z^{10}. 
\end{eqnarray}
We can also compute the radius of convergence of the virial expansion.
A simple analysis of the integral (\ref{ON-int}) shows that the singularity 
in the complex $c$-plane that is closest to the origin corresponds to 
$c u n^2 = -1$. This implies $A_n/A_{n-1} = -8 \pi^{3/2} z$ asymptotically, and 
that the virial expansion converges for
$|\Phi_p| < 1/(8 \pi^{3/2} z)$ in the limit $z\to 0$.

Using the expressions reported in Ref.~\CITE{ON-83} we obtain
for the radius of gyration:
\begin{eqnarray}
S_{1,g}(z) &=& {64\over 3465} (1365 - 1028 \sqrt{2}) \pi^{3/2} z^2 = 
   -9.13421 z^2, \\
S_{2,g}(z) &=& {2048\over 405405} 
           (85013 - 115408 \sqrt{2} + 45684 \sqrt{3}) \pi^3  z^3 = 
   145.428 z^3, 
\label{S2g-z}\\
S_{3,g}(z) &=& -3113.85 z^4, \\
S_{4,g}(z) &=& 78503.4 z^5, \\
S_{5,g}(z) &=& -2.19663\cdot 10^6\, z^6 .
\end{eqnarray}
For the end-to-end distance we obtain analogously:
\begin{eqnarray}
S_{1,e}(z) &=& {128\over 315} (103 - 76 \sqrt{2}) \pi^{3/2} z^2 = 
   -10.1374 z^2, \\
S_{2,e}(z) &=& {4096\over 10395} 
           (1427 - 3248\sqrt{2} + 1836 \sqrt{3}) \pi^3  z^3 = 
   167.132 z^3, 
\label{S2e-z}\\
S_{3,e}(z) &=& -3654.20 z^4, \\
S_{4,e}(z) &=& 93359.1 z^5 , \\
S_{5,e}(z) &=& -2.6358\cdot 10^6\,  z^6 .
\end{eqnarray}
Finally, for the hydrodynamic radius we use the representation
\be
{1\over R_H} = 4 \pi \int {d^3 q\over (2\pi)^3} {F(q)\over q^2},
\ee
where $F(q)$ is the form factor 
\be
F(q) = {1\over N^2} \left\langle \sum_{ij}
   e^{i\mathbf{q} \cdot (\mathbf{r}_i - \mathbf{r}_j)}\right\rangle,
\ee
which is normalized so that $F(q) = 1 - {q^2\over3} R^2_g  + O(q^4)$.
For the density corrections, we obtain at leading order
\begin{eqnarray}
S_{1,H}(z) &=&  3.32047z^2 + O(z^3) \; , \\
S_{2,H}(z) &=& - 44.8425z^3 + O(z^4) \; .
\end{eqnarray}
Finally, we report the swelling factors:
\begin{eqnarray}
\alpha_e^2 &=& 1 + {4\over 3} z + O(z^2), \\
\alpha_g^2 &=& 1 + {134\over 105} z + O(z^2), \\
\alpha_H &=& 1 -  \left({27\pi\over 16} - 4 - {3\pi\over2}\log {3\over2}
       \right) z + O(z^2) \; .
\end{eqnarray}
Additional terms for $\alpha^2_e$ are reported in Ref.~\CITE{CCPRV-01}.

\clearpage

\begin{table}
\caption{Estimates of $A_2(z)$ for five different values of $z$ and different 
$N_{\rm min}$.}
\label{tab-A2}
\begin{tabular}{ccccc}
$z$ & $N_{\rm min}=100$ & $N_{\rm min}=250$ & $N_{\rm min}=500$ &
$N_{\rm min}=1000$  \\
\hline
$z_1=0.056215$ & 0.99241(41) &  0.99212(59) &  0.99257(98) & 0.9906(19) \\
$z_2=0.148726$ & 1.97964(79) &1.9796(12)  & 1.9782(18)  &1.9831(35) \\
$z_3=0.321650$ & 2.9646(11) & 2.9649(16) &2.9621(27) & 2.9642(48)\\
$z_4=0.728877$ & 3.9469(14) & 3.9452(21) & 3.9433(34) & 3.9498(65)\\
$z_5=2.508280$ & 4.9264(15) & 4.9221(23) &4.9147(36) & 4.9169(65)\\
\end{tabular}
\end{table}

\begin{table}
\caption{TPM estimates of several quantities for five different values of $z$.
$z_1$, $z_2$, $z_3$, $z_4$, and $z_5$ are reported in the text.
}
\label{table-TPM}
\begin{tabular}{cccccc}
      & $z_1$ &  $z_2$ &  $z_3$ &  $z_4$ &  $z_5$ \\
\hline
$A_2$  & 0.99257(98) & 1.9782(18) &2.9621(27) &3.9433(34) &4.9147(36) \\
$A_3$  &0.08486(78) &0.6061(30) &1.8435(82) &4.021(13) &7.243(22) \\
$A_4$  &$-$0.095(11) &$-$0.857(78) &$-$2.77(30) &$-$5.51(79) &$-$10.3(1.8) \\
$A_{ge}$&0.166145(65) &0.165517(62) &0.164591(65) &0.163272(62) &0.161514(55) \\$A_{gH}$  &1.50553(84) &1.50964(90) &1.51731(78) & 1.52597(74) & 1.54165(69)\\
$S_{1,g}$  &$-$0.01711(66) &$-$0.0630(11) &$-$0.1278(18) &$-$0.1970(24) &$-$0.2747(25) \\
$S_{2,g}$  &0.0073(12) &0.0309(42) &0.051(10) &0.082(17) &$-$0.002(25) \\
$S_{1,e}$  &$-$0.01931(71) &$-$0.0718(13) &$-$0.1469(20) &$-$0.2340(27) &$-$0.3301(30) \\
$S_{2,e}$  &0.0097(14) &0.0381(53) & 0.083(12) &0.124(21) & 0.035(33)\\
$S_{1,H}$  &0.00527(36) &0.02079(75) &0.0405(11) &0.0602(13) &0.0771(15) \\
$S_{2,H}$  &$-$0.00197(84) &$-$0.0086(33) &$-$0.0041(66) &0.026(12) &0.028(17) \\
$\alpha_e^2$  &1.06872(81) &1.16553(82) &1.31167(92) &1.5683(12) &2.2283(15) \\
$\alpha_g^2$  &1.06545(71) &1.15751(78) &1.29540(89) &1.5362(11) &2.1595(13) \\
$\alpha_H$  &1.03127(44) &1.07196(44) &1.12878(42) &1.22196(40) &1.43419(43) \\
\end{tabular}
\end{table}

\begin{table}
\caption{TPM interpolation formulas.}
\label{table-interp}
\begin{tabular}{lc}
$A_2(z)$ & $4 \pi^{3/2} z 
   (1 + 19.1187 z + 126.783 z^2 + 331.99 z^3 + 268.96 z^4)^{-1/4}$\\
$A_3(z)$ & $1100.7 z^3
   (1 + 23.1258 z + 195.358 z^2 + 594.386 z^3 + 541.906 z^4)^{-3/4}$ \\ 
$A_4(z)$ & $-29883.1 z^4
   (1 + 17.4354 z + 135.853 z^2 + 437.409 z^3)^{-4/3}$ \\
$A_{ge}(z)$ & $1/6 - 0.0571429 z
   (1 + 1455.41 z + 47738.2 z^2  + 584.595 z^3  + 5025.99 z^4)^{-1/4}$ \\
$A_{gH}(z)$ & $1.50451 + 0.0288252 z
   (1 - 1.64436 z + 1.57777 z^2  + 0.0535093 z^3)^{-1/3}$ \\
$S_{1,g}(z)$ & $-9.13421 (1 + 0.0182093 z) z^2/
   (1 + 10.8944 z + 28.8313 z^2  + 0.52769 z^3)$ \\
$S_{2,g}(z)$ & $145.428 (1 - 0.39868 z) z^3 
   (1 + 25.0806 z + 92.4415 z^2  + 131.164 z^3)^{-4/3}$ \\
$S_{1,e}(z)$ & $-10.1374 (1 + 0.0164091 z) z^2/
   (1 + 10.8171 z + 26.1977 z^2  + 0.431731 z^3)$ \\
$S_{2,e}(z)$ & $167.132 (1 - 0.161819 z) z^3
   (1 + 24.6508 z + 52.9313 z^2  + 178.051 z^3)^{-4/3}$ \\
$S_{1,H}(z)$ & $3.32047 z^2 
   (1 + 23.196 z + 71.9449 z^2  + 257.679 z^3)^{-2/3}$ \\
$S_{2,H}(z)$ & $-44.8425 (1 - 2.7513z) z^3
   (1 + 28.5327 z - 37.0818 z^2  + 350.101 z^3)^{-4/3}$ \\
$\alpha_g(z)$ & 
    $(1 + 10.9288 z + 35.1869 z^2  + 30.4463 z^3)^{0.0583867}$ \\
$\alpha_e(z)$ & 
    $(1 + 11.4181 z + 39.7661 z^2  + 42.8257 z^3)^{0.0583867}$ \\
$\alpha_H(z)$ & $(1 + 10.4351 z + 29.7693 z^2  + 16.8909 z^3 )^{0.0583867}$ \\
\end{tabular}
\end{table}

\clearpage

\clearpage

\begin{thebibliography}{199}
\bibitem{footPsi}
In experimental work the low-density behavior of
the osmotic pressure $\Pi$ is usually written as 
$\Pi/(R T\rho) = 1/M + B_2 \rho + O(\rho^2)$, where
$M$ is
the molar mass of the polymer, $\rho$ the weight concentration, and
$T$ the absolute temperature. Then, 
$\Psi \equiv 2 (4\pi)^{-3/2} M^2 B_2 \hat{R}^{-3}_g/N_A$, where $N_A$ is the
Avogadro number.
\bibitem{footnote-bn}
To prove that $a_n/a_2 = b_n$, we write $A_n = A_n^* + A_{1,n}(w) N^{-\Delta}$.
For $w \to 0$, consistency with Eq.~(\ref{Anz-exp}) requires 
$A_{1,n}(w) =  a_n (k w)^{-2\Delta}$, where $k$ is defined by 
$z = k w N^{1/2}$. Since $A_{1,n}(w)/A_{1,2}(w) = b_n$ for any $n$, 
we obtain $a_n/a_2 = b_n$.
\bibitem{foot-A2-Sch}
We relate the constant $\tilde{z}$ of Ref.~\CITE{Schaefer-99} to $z$
by requiring $A_2(z) \approx {1\over2} (4\pi)^{3/2} z$ for $z\to 0$:
we obtain $z = 0.182 \tilde{z}$.
\bibitem{footnoteLue}
The compressibility factor $Z$ in Refs.~\CITE{LK-02,Lue-00} is expressed
in terms of $X\equiv B_2 c$, where $B_2$ is the second virial coefficient
defined by $Z = 1 + B_2 c + O(c^2)$. In order to relate $X$ to $\Phi_p$
we note that $X = A_2 c \hat{R}_g^3 = A_2^* (\Psi/\Psi^*) 3 \Phi_p/(4 \pi)$
and use $A_2^* = 5.500$ (Ref.~\CITE{CMP-vir}).
\bibitem{footnote_c0} 
It is possible to riduce the discrepancy between field theory 
and our results by changing the parameters $c_0$ and $n_0$. 
For the values suggested in Ref.~\CITE{Schaefer-99}, $c_0 = 1.2$ and 
$n_0=0.53$, $R^2_g/\hat{R}_g^2 = 0.857$ for $\Phi_p = 1$,
to be compared with our result
$0.920$. The discrepancy decreases if $c_0$ is increased:
for $n_0 = 0.53$ and $c_0 = 6.9$ field theory gives the same result as ours. 
However, $c_0$ cannot be too large, otherwise $Z$ is underestimated in the 
semidilute regime. For $c_0 = 6.9$, field theory predicts 
$Z = 1.20 \Phi_p^{1.309}$, while experiments, other field-theory 
calculations, and the extrapolation of the virial expansion predict 
$Z = a \Phi_p^{1.31}$ with $a = 1.5$-1.7 (see the conclusions of 
Ref.~\CITE{CMP-vir} for a summary of the existing results).
\end{thebibliography}
\end{document}